\documentclass{aastex631}

\newcommand\aastex{AAS\TeX}

\received{July 28, 2022}
\revised{January 23, 2023}
\accepted{February 16, 2023}

\submitjournal{ApJ}

\usepackage{amsmath}
\usepackage{url}
\usepackage{comment}

\usepackage{ulem}

\defcitealias{Ikuta20}{Paper~I}
\graphicspath{{./}{figures/}}

\shorttitle{\aastex\ draft}
\shortauthors{Ikuta et al.}

\renewcommand{\baselinestretch}{1.5}

\begin{document}

\title{Starspot mapping with adaptive parallel tempering. I\hspace{-.1em}I. Application to TESS data for M-dwarf flare stars, AU Microscopii, YZ Canis Minoris, and EV Lacertae
}

\correspondingauthor{Kai Ikuta}
\email{ikuta@kusastro.kyoto-u.ac.jp}

\author[0000-0002-5978-057X]{Kai Ikuta}
\altaffiliation{Current Affiliation: Department of Multidisciplinary Sciences, The University of Tokyo, 3-8-1 Komaba, Meguro, Tokyo 153-8902, Japan}
\affil{Astronomical Observatory, Kyoto University, Kitashirakawa-Oiwake-cho, Sakyo, Kyoto 606-8502, Japan}

\author[0000-0002-1297-9485]{Kosuke Namekata}
\affil{ALMA Project, National Astronomical Observatory of Japan, NINS, Mitaka, Tokyo 181-8588, Japan}

\author[0000-0002-0412-0849]{Yuta Notsu}
\affil{Department of Earth and Planetary Sciences, Tokyo Institute of Technology, 2-12-1 Ookayama, Meguro, Tokyo 152-8551, Japan}
\affil{Laboratory for Atmospheric and Space Physics, University of Colorado Boulder, 3665 Discovery Drive, Boulder, Colorado 80303, USA}
\affil{National Solar Observatory, 3665 Discovery Drive, Boulder, Colorado 80303, USA}

\author[0000-0003-0332-0811]{Hiroyuki Maehara}
\affil{Astronomical Observatory, Kyoto University, Kitashirakawa-Oiwake-cho, Sakyo, Kyoto 606-8502, Japan}
\affil{Okayama Branch Office, Subaru Telescope, National Astronomical Observatory of Japan, NINS, Kamogata, Asakuchi, Okayama 719-0232, Japan}

\author{Soshi Okamoto}
\affil{Department of Astronomy, Kyoto University, Kitashirakawa-Oiwake-cho, Sakyo, Kyoto 606-8502, Japan}
\affil{Japan Meteorological Agency, 1-3-4 Otemachi, Chiyoda, Tokyo 100-8122, Japan}

\author{Satoshi Honda}
\affil{Nishi-Harima Astronomical Observatory, University of Hyogo, 407-2 Nishigaichi, Sayo-cho, Sayo, Hyogo 679-5313, Japan}

\author{Daisaku Nogami}
\affil{Department of Astronomy, Kyoto University, Kitashirakawa-Oiwake-cho, Sakyo, Kyoto 606-8502, Japan}

\author{Kazunari Shibata}
\affil{Kwasan Observatory, Kyoto University, 17 Ohmine-cho, Kita-Kazan, Yamashina, Kyoto 607-8471, Japan}
\affil{Department of Environmental Systems Science, Doshisha University, 1-3 Tataramiyakodani, Kyotanabe, Kyoto 610-0394, Japan}

\begin{abstract} 
Starspots and stellar flares are indicators of stellar magnetic activity.
The magnetic energy stored around spots is thought to be the origin of flares, but the connection is not completely understood. To investigate the relation between spot locations deduced from the light curves and occurrence of flares therein, we perform starspot modeling for TESS light curves of three M-dwarf flare stars, AU Mic, YZ CMi, and EV Lac, using the code implemented in Paper I. The code enables to deduce multiple stellar/spot parameters by the adaptive parallel tempering algorithm efficiently. We found that flare occurrence frequency is not necessarily correlated with the rotation phases of the light curve for each star. The result of starspot modeling shows that either spot is always visible to the line of sight in all phases, and we suggest that this can be one of the reasons that there is no or less correlation between rotation phases and flare frequency. In addition, the amplitude and shape of the light curve for AU Mic and YZ CMi have varied in two years between different TESS Cycles. The result of starspot modeling suggests that this can be explained by the variations of spot size and latitude.
\end{abstract}


\section{Introduction} \label{sec:intro}
Starspots are apparent manifestations of magnetic activity on the stellar surface, and can be ubiquitously observed on various type of stars \citep[for reviews,][]{2005LRSP....2....8B,2009A&ARv..17..251S,2021A&ARv..29....1K}. For active young stars, cool stars (M, K, G-dwarfs), and RS CVn-type stars, starspots have been extensively studied through ground-based observations of photometry \citep[e.g.,][]{1995ApJS...97..513H,2002A&A...393..225M} and spectroscopy \citep[e.g.,][]{1992A&A...259..183S,1999A&A...347..212S}.
Stellar flares are intense explosions in the stellar atmosphere by releasing magnetic energy around starspots, and have extensively been studied in close association with spots \citep[e.g.,][]{1980AJ.....85..871P,1991ApJ...378..725H,2005stam.book.....G,2010ARA&A..48..241B}. In particular, starspot and stellar flare studies have been enabled with high precision and long-term photometry by the advent of Kepler space telescope \citep{2010ApJ...713L..79K} and Transiting Exoplanet Survey Satellite \citep[TESS;][]{2015JATIS...1a4003R}.
Many superflares are reported on a large number of cool stars for 
Kepler data \citep{Maehara12, 2014ApJ...797..121H, 2014ApJ...797..122D, 2016NatCo...711058K, 2019ApJ...876...58N,Okamoto20} and TESS data \citep[e.g.,][]{2020AJ....159...60G, 2020ApJ...890...46T, 2020AJ....160..219F, 2021ApJS..253...35T,2022NatAs...6..241N,2022ApJ...926L...5N}.
The spot size estimated from the amplitude of the light curve is determined by the upper limit of the flare energy for Kepler data of solar-type stars, and stars with larger spots cause larger flares as in the case of solar flares \citep{2013ApJ...771..127N,2013PASJ...65...49S,2017PASJ...69...41M,Okamoto20}.
\cite{2014ApJ...797..121H} and \cite{2016ApJ...829..129S} investigate the relation between the rotation phase of Kepler light curve and occurrence of flares therein for an active dMe star GJ1243 and found that the rotation phase and flare frequency are not correlated. They suggest that this could be because there are large spots on the visible pole or small spots distribute in the whole stellar disk.
The ensemble studies also suggest that the flare frequency is not necessarily correlated with the phase of the light curve for Kepler/K2 data \citep[e.g.,][]{2018MNRAS.480.2153D} and TESS data \citep[][]{2020MNRAS.494.3596D,2020AJ....160..219F}.

It is unknown where the spots are actually on the stellar surface because the surface cannot be spatially resolved with a simple analysis from the light curve.
Therefore, it is important to investigate the relation between the spot locations and flare occurrence by mapping the surface with sophisticated analysis, hereafter referred to as starspot modeling.
\cite{2015ApJ...806..212D} conducted starspot modeling for the Kepler light curve of GJ1243 with two-spot model and showed that the light curve can be explained by a spot on the stellar equator and the other spot at the high latitude. Their starspot modeling supports the suggestion by \cite{2014ApJ...797..121H} that flare frequency is not correlated with the spot location because of a always visible large spot on the pole.
This kind of sophisticated analysis can be important to understand the origin of flares, but a small number of stars including GJ1243 have been investigated from this point. Then, it is also necessary for modeling of more number of light curves to understand the relation between spot locations and flare frequency on flare stars.

AU Microscopii (AU Mic), YZ Canis Minoris (YZ CMi), and EV Lacertae (EV Lac) are bright and magnetically active M-dwarf stars. These have been observed for monitoring their flares and to investigate their spots by (Zeeman) Doppler imaging through ground-based observations. Recently, these targets have been spectroscopically observed at the same time as the TESS observations to delve into the flare mechanisms \citep[e.g.,][]{Maehara2020}. The spot locations can be one of indicators where flares occur on the stellar disk and which direction the plasma associated with flares erupts to \citep[][]{2022NatAs...6..241N}.
AU Mic is a young dM1e dwarf with a spatially resolved debris disk \citep{2004Sci...303.1990K}, and was revealed to be a young planetary system harboring two warm Neptunes \citep{2020Natur.582..497P, 2020A&A...641L...1M,2022AJ....163..147G} from the transit search \citep[][]{2007MNRAS.379...63H}.
Since AU Mic is magnetically active and frequently causes large flares \citep{1970PASP...82.1341K}, it has been spectroscopically observed from radio \citep {1987ApJ...312..822K} to X-ray \citep{2005A&A...431..679M}.  Its surface is largely filled with strong magnetic fields \citep{Klein20,2022MNRAS.512.5067K}, and the light curve exhibits rotational periodic modulations ascribed to starspots on the surface \citep{1972ApL....11...13T,1986A&A...165..135R, 1987A&A...174..139B, 2020Natur.582..497P}.
YZ CMi is a magnetically active dM4.5e dwarf.
Flares are reported from radio \citep{1974ApJ...190L.129S} through optical \citep{1976ApJS...30...85L} to X-ray \citep{1975ICRC....1..154G}, and has been spectroscopically observed for the flare properties \citep{ 2010ApJ...714L..98K,Maehara2020}.
The modulation of light curve is reported through ground-based observations \citep{1983A&A...123..184P,2003AN....324..527Z}. There are spots on the pole by Zeeman Doppler imaging \citep{2008MNRAS.390..567M}.
EV Lac is a magnetically active dM3.5e dwarf and has been investigated for their flare properties \citep{2005ApJ...621..398O,2018PASJ...70...62H, 2020MNRAS.499.5047M,2021ApJ...922...31P}. The modulation of light curve is reported through ground-based observations \citep{1980AJ.....85..871P,1983A&A...123..184P}, and there are also spots on the pole by (Zeeman) Doppler imaging \citep{2008MNRAS.390..567M,2022arXiv220300415J}.

In this study, for the purpose of investigating the spot properties and their relation with flare occurrences, we conduct starspot modeling for TESS light curves of M-dwarf flare stars, AU Mic, YZ CMi, and EV Lac, using a code implemented in \cite{Ikuta20} (hereafter, referred to as \citetalias{Ikuta20}). 
In \citetalias{Ikuta20}, the code is implemented for starspot modeling with an adaptive parallel tempering (PT) algorithm
\citep[e.g.,][]{1999PhRvE..60.3606H,2016MNRAS.455.1919V} and an importance sampling algorithm to deduce stellar/spot parameters specified in \texttt{macula} \citep{2012MNRAS.427.2487K} and compare the number of spots in the Bayesian framework. %To evaluate the performance of the code, we apply the code to synthetic light curves emulating Kepler data. 
This paper is organized as follows. In Section \ref{sec:method}, we describe the TESS data, flare detection, calculation of flare energy, and the numerical setup for starspot modeling of the data. In Section \ref{sec:result}, we discuss the results of starspot modeling in terms of comparing with previous studies by photometry and spectroscopy, the relation between spot locations and flare frequency, the variation of the amplitude and shape of the light curve for AU Mic and YZ CMi in two years, and comparing the number of spots in the Bayesian framework in addition to refering the validity of starspot modeling. In Section \ref{sec:summary}, we conclude this paper and describe future prospects. {In Appendix \ref{sec:appendixb}, we describe the result with different assumptions for the light curve of EV Lac Cycle 2 to evaluate the uncertainties of parameters.}
In Appendix \ref{sec:appendix}, we exhibit supplementary figures delineating the joint posterior distributions for each of the model.

\section{Method} \label{sec:method}
\subsection{Preprocessing TESS data and flare detection} \label{sec:dataset}
AU Mic (TIC 441420236), YZ CMi (TIC 266744225), and EV Lac (TIC 154101678) were observed with 2-min cadence by TESS in Sector 1 (July 25 to August 22 in 2018), 7 (January 7 to February 2 in 2019), and 16 (September 11 to October 7 in 2019) through the TESS prime mission (Cycle 1 and 2), respectively.
After that, AU Mic and YZ CMi were also observed in Sector 27 (July 4 to 30 in 2020) and 34 (January 13 to February 9 in 2021) through the TESS extended mission (Cycle 3), respectively.
We retrieve the PDC-SAP (Pre-search Data Conditioning Simple Aperture Photometry) flux from the MAST (Mikulski Archive for Space Telescopes) Portal webcite.\footnote{\url{https://mast.stsci.edu/portal/Mashup/Clients/Mast/Portal.html}}
{To confirm effects by the TESS systematics due to scattered light and spacecraft drift, we derive the rotation period from the SAP light curves using the TESS Systematics-insensitive Periodogram \citep{2020RNAAS...4..220H}. As a result, we obtain the periods of $4.872$, $4.825$, $2.778$, $2.778$, and $4.364$ days, for AU Mic Cycle 1 and 3, YZ CMi Cycle 1 and 3, and EV Lac Cycle 2, respectively. These values are consistent with the results deduced from the PDC-SAP light curves, and the detrended light curves correspond to that of the PDC-SAP light curves. Thus, we adopt the PDC-SAP light curves for the starspot modeling. For the case of EV Lac Cycle 2 as an example case, we exhibit the result by the two-spot model with the systematic noise model by Gaussian process (Appendix \ref{sec:a3}).}

Figure \ref{fig:lc} exhibits TESS light curves in terms of the relative flux (black), the PDC-SAP flux normalized by its average excluding missing values. Vertical lines in each figure also represent the detected flares at each time (red) as following procedures.
First, we detect flare candidates in the light curves 
by using \texttt{stella}
\citep[][]{2020JOSS....5.2347F}. The code detects flares from TESS light curves with convolutional neural networks, 
and we adopt their pretrained model with detected flares in \cite{2020AJ....159...60G}.
Then, we adopt the flare probability of 0.1 and an amplitude of $10^{-3}$ for all trained hundred models as fewer false negatives as possible. We confirm the flare candidates as flares by eye because many false positives are included in the detected candidates \citep[][]{2022ApJ...925L...9F}.
Second, after extracting confirmed flares from the light curves, we also extract outliers of the modulations ascribed to spots from the remaining light curves with Bessel filter \citep[][]{Maehara2020}.
Although a small number of outliers remain in the light curves, they do not affect the result of the starspot modeling since the number of data points is much less than that of total data points.
In this study, since the modulation timescale ascribed to spots is much longer than the cadence, we use data every three points for computational efficiency.
As a result of the flare detection, the number of detected flares are summed up to 55, 70, 124, 137, and 76, for AU Mic Cycle 1 and 3, YZ CMi Cycle 1 and 3, and EV Lac Cycle 2, respectively (Figure \ref{fig:lc}).\footnote{Each flare properties is listed on \url{https://github.com/KaiIkuta/StarspotMapping.git}: the peak time (BJD-2745000), amplitude, equivalent duration (sec), and e-folding time (day).}
The detectability of flares may depends on the rotation phase of the light curve because the amplitude of the light curve for these stars are large, and flares in the training set of the pretrained model are detected by reducing the periodic signal of the light curve \citep{2020AJ....159...60G}. Thus, we consider the equivalent duration = 1 (sec) as the sufficient threshold to discuss the correlation between spot locations and flare frequency.
Among detected flares, the number of flares with larger equivalent duration than 1 (sec) are 48, 50, 109, 106, and 71, respectively.

According to \cite{2013ApJS..209....5S}, a bolometric flare energy is formulated as

\begin{align}
E_{\rm flare} &= \int L_{\rm flare} (t) dt %\notag 
= \sigma_{\rm SB} T_{\rm flare}^4 \times  2 \pi R_{\rm star}^2 \times \frac{\int R(\lambda) B(\lambda,T_{\rm eff}) d\lambda}{\int R(\lambda) B(\lambda,T_{\rm flare}) d\lambda} \times \int \frac{\Delta F}{F} (t) dt,
\end{align}
where $L_{\rm flare} (t)$ and $\Delta F/F (t)$ are the bolometric luminosity and relative flux at the time $t$, and $\sigma_{\rm SB}= 5.67\times 10^{-1}$ (erg K$^{-4}$ m$^{-2}$ sec$^{-1}$) is the Stefan–Boltzmann constant.
The flare is assumed to be the blackbody radiation with the flare temperature $T_{\rm flare}=10000$ (K).
The last time-integrated factor corresponds to the equivalent duration, which is calculated by the integral of the relative flux with time from the start to end of a flare \citep{2014ApJ...797..121H}. The coefficient specific to a star is the flare energy with the equivalent duration $=1$ (sec).
Those for AU Mic, YZ CMi, and EV Lac are calculated to be $3.68 \times 10^{32}$, $3.56 \times 10^{31}$, $5.25 \times 10^{31}$ (erg), respectively.
We note that the calculation error up to a factor is inherent in the assumed flare model \citep{Maehara2020}.
If assuming the flare temperature $T_{\rm flare}=7000$ (K), the energy reduces by a factor of two.
Figure \ref{fig:freq1}, \ref{fig:freq2}, and \ref{fig:freq3} accumulates flares as their flare frequency for AU Mic, YZ CMi, and EV Lac, respectively \citep[cf.][]{Maehara2020}. We discuss the relation between spot locations and flare frequency in Section \ref{sec:phase} and \ref{sec:str}.

\subsection{Numerical setup} \label{sec:set}
Table \ref{tb:para0} lists stellar parameters, references, deduced spot temperature and intensity, and limb-darkening coefficients for AU Mic, YZ CMi, and EV Lac.
The spot temperature $T_{\rm spot}$ is deduced from the quadratic formula in the stellar effective temperature $T_{\rm eff}$ \citep[Equation 4 in][]{2021ApJ...907...89H}:
\begin{equation} \label{eq:spottemp}
T_{\rm spot} = -3.58 \times 10^{-5} T_{\rm eff}^2 + 0.801 T_{\rm eff}+666.5 .
\end{equation}
Thereby, the spot relative intensity $f_{\rm spot}$ for TESS band is calculated by the stellar effective temperature $T_{\rm eff}$ and the spot temperature $T_{\rm spot}$:
\begin{equation} \label{eq:relspot} 
f_{\rm spot} = \frac{\int R(\lambda) B(\lambda,T_{\rm spot}) d\lambda}{\int R(\lambda) B(\lambda,T_{\rm eff}) d\lambda},
\end{equation}
where $R(\lambda)$ and $B(\lambda,T)$ are the TESS-band response function and the Plank function with each wavelength $\lambda$, respectively.
The stellar limb-darkening law is adopted as the following formula for the cosine of the angle to the line of sight $\mu$: 
\begin{equation} \label{eq:limb}
I(\mu)/I(1)= 1 - \Sigma_{k=1}^4 c_k(1-\mu^{k/2}),
\end{equation}
and the coefficients $c_k$ based on the stellar parameters are deduced from \cite{Claret18}. {Then, we adopt the nonlinear limb-darkening law. For the case of EV Lac Cycle 2 as an example case, we exhibit the result by the quadratic limb-darkening law (Appendix \ref{sec:a}).}

In \citetalias{Ikuta20}, we implement a code for starspot modeling using an adaptive PT algorithm. The model evidence for each model is also computed with an importance sampling algorithm together with the PT transition. We adopt an analytical spotted model \citep[\texttt{macula};][]{2012MNRAS.427.2487K} for the spot model, and the model is specified in the stellar and spot parameters: the stellar inclination angle $i$ (deg); equatorial rotation period $P_{\rm eq}$ (day); degree of differential rotation $\kappa$; spot relative intensity $f_{\rm spot}$; latitude $\Phi$ (deg); initial longitude $\Lambda$ (deg); reference time  $t_{\rm ref}$ (day) (the time at the midpoint of the interval over
which the spot has its maximum radius); maximum radius $\alpha_{\rm max}$ (deg); emergence duration ${\cal I}$ (day); decay duration ${\cal E}$ (day);
stable duration ${\cal L}$ (day).
The rotation period at the latitude $\Phi$  (Equation 16 in \citetalias{Ikuta20}) is described by
\begin{equation} \label{eq:rot}
P(\Phi)=\frac{P_{\rm eq}}{1-\kappa \sin^2 \Phi}.
\end{equation}

In this study, it is unnecessary to adopt the spot parameters of the reference time $t_p$, emergence duration ${\cal I}$, decay duration ${\cal E}$, and stable duration ${\cal L}$ in the model because the amplitudes of the light curves are approximately constant in the observation period of a TESS Sector. Therefore, the light curves are specified with fewer stellar and spot parameters than those used in \citetalias{Ikuta20} for simplicity (Table \ref{tb:para1f}, \ref{tb:para1s}, \ref{tb:para2f}, \ref{tb:para2s}, and \ref{tb:para3f}). In fact, we ascertain that the posterior distribution does not converge to an unimodal distribution due to these spot parameters by exploiting the test runs.
Each of the spot is discerned by the range of latitude $\Phi$ from the southern to northern pole, unlike the range of the reference time $t_p$ in \citetalias{Ikuta20}.

Light curves of AU Mic Cycle 1 and 3, YZ CMi Cycle 3, and EV Lac exhibit two local minimum per one equatorial rotation period, and we optimize them by two- and three-spot models (Table \ref{tb:para1f}, \ref{tb:para1s}, \ref{tb:para2s}, and \ref{tb:para3f}).
The light curve of YZ CMi Cycle 1 exhibits one local minimum per one equatorial rotation period, and we optimize it by two- and one-spot models (Table \ref{tb:para2f}). As for the one-spot model, the differential rotation parameter $\kappa$ is fixed to 0. We note that the light curves of YZ CMi have one local minimum per one equatorial rotation period possibly due to its low inclination angle \citep[e.g.,][]{Basri20}.
We ascertained the posterior distributions of the inclination angle and spot intensity {would} converge to unphysical modes (the edges of the parameter range) by exploiting test runs because they have respective degeneracies with the spot latitude and radius \citepalias{Ikuta20}. Therefore, the values of the inclination angle and spot intensity are fixed to the values in Table \ref{tb:para0}, and we abbreviate the name followed by \citetalias{Ikuta20} for each of the model: for instance, ``two-spot model fixed $\sin i$ and $f_{\rm spot}$'' in the context of \citetalias{Ikuta20} is simply renamed ``two-spot model'' in this study. {For the case of EV Lac Cycle 2 as an example case, we exhibit the result that the spot intensity is included in the parameters under an uniform and normal prior distributions (Appendix \ref{sec:ab}).}

\begin{deluxetable*}{lccc}
\tablecaption{Stellar parameters\label{tb:para0}}
\tabletypesize{\footnotesize}
\tablehead{
\colhead{Stellar parameter} & \colhead{AU Mic} & \colhead{YZ CMi} &
\colhead{EV Lac}}
\startdata
Effective temperature $T_{{\rm eff}}$ (K)&$3700 \pm 100$ $^c$ & $3100\pm 50$ $^e$ & $3400\pm18$ $^h$  \\
Rotation period $P_{{\rm rot}}$ (day)&  $4.863\pm 0.010$ $^c$&$2.776\pm 0.010$ $^e$ & $4.359$ $^j$ \\
Stellar radius $R_{\rm star}$ ($R_{\rm sun}$)  &$0.75\pm 0.03$ $^c$&  $0.37^{+0.03}_{-0.06}$ $^e$ & $0.35\pm 0.02$ $^j$  \\
Surface gravity $\log g$\tablenotemark{a}& $4.39\pm 0.10$ &$4.87 \pm 0.33$ &  $4.89\pm0.13$ \\
Inclination angle $i$ (deg)&$75$ $^d$&$36^{+17}_{-14}$ $^e$ & $60$ $^k$ \\ \hline
Spot temperature $T_{\rm spot}$ (K)\tablenotemark{a} &  $3140\pm 54$ &  $2806\pm 29$  &  $2976\pm 10$ \\
Spot relative intensity  $f_{\rm spot}$\tablenotemark{a}&$0.43\pm 0.04$&$0.56\pm 0.03$&$0.48\pm 0.01$ \\ \hline
Limb-darkening coefficients ($c_1$, $c_2$, $c_3$, $c_4$)\tablenotemark{b} & ($2.87$, $-4.35$, $3.92$, $-1.32$)& ($3.15$, $-4.62$, $4.01$, $-1.34$) & ($3.00$,$-4.54$, $4.01$, $-1.35$) \\
\enddata
\tablenotetext{a}{
{The surface gravity is calculated from the stellar radius $R_{\rm star}$ and mass $M_{\rm star}$.} The spot temperature is formulated by the stellar effective temperature $T_{{\rm eff}}$ in Equation \ref{eq:spottemp}, and the spot relative intensity $f_{\rm spot}$ for TESS band is deduced from Equation \ref{eq:relspot}.
}
\tablenotetext{b}{
The stellar limb-darkening coefficient for TESS data is characterized by the effective temperature $T_{\rm eff}$ and surface gravity $\log g$ under the solar metalicity \citep[][]{Claret18}. {We adopt the coefficient values of ($T_{\rm eff}$, $\log g$)=($3700$, $4.5$), ($3100$, $5.0$), and ($3400$, $5.0$), for AU Mic, YZ CMi, and EV Lac, respectively.}
}
\tablerefs{}{$^c$ \cite{2020Natur.582..497P};$^d$ \cite{2019ApJ...883L...8W}; $^e$ \cite{Baroch20}; $^f$ \cite{Cifuentes20}; $^h$ \cite{2012ApJ...748...93R}; $^j$ \cite{2021ApJ...922...31P}; $^k$ \cite{2008MNRAS.390..567M};}
\end{deluxetable*}

\section{Results and discussion} \label{sec:result}
We optimize the light curves of AU Mic Cycle 1 and 3, YZ CMi Cycle 3, and EV Lac Cycle 2, by the two- and three-spot models, and the light curve of YZ CMi Cycle 1 by the two- and one-spot models.
The number of spots is set as the number of local minimum in the light curve and one more number (Section \ref{sec:set}).
For every cases, unimodal posterior distributions are deduced regardless of the number of spots, and joint ones are delineated in Appendix \ref{sec:appendix}. 
Table \ref{tb:para1f}, \ref{tb:para1s}, \ref{tb:para2f}, \ref{tb:para2s}, and \ref{tb:para3f} show the modes of the deduced posterior distributions, their credible regions, reproduced average flux, and the {logarithm of} model evidence for each model, together with their prior distributions for each of the parameters, {for AU Mic Cycle 1 and 3, YZ CMi Cycle 1 and 3, and EV Lac Cycle 2,} respectively. 
{The preferable models and their values of the} {logarithm of} {model evidence are marked in bold. Figure \ref{fig:lc} exhibits the reproduced light curves with each mode of the posterior distributions in each of the two-spot model (red).}
%Figure \ref{fig:plot1f}, \ref{fig:plot1s}, \ref{fig:plot2f}, \ref{fig:plot2s}, and \ref{fig:plot3f} exhibit the light curves (gray), reproduced ones with each mode of the posterior distributions (red), and their residual (black). 
All light curves can be folded in phase with the rotation period
since the deduced degree of differential rotation $\kappa$ is approximately 0 or the absolute value of spot latitudes is almost the same value, and the amplitude of the light curve is approximately constant in the observation period of a TESS Sector.
Figure \ref{fig:phase1f}, \ref{fig:phase1s}, \ref{fig:phase2f}, \ref{fig:phase2s}, and \ref{fig:phase3f} exhibit visualized surface reproduced with each mode of posterior distributions; (a, f) phase-folded light curve (black) and reproduced one with the model (red); 
 (b, g) the temporal variation of visible projected area of each spot (red, blue, and green) and the total (black) relative to the disk; (c, h) that of visible area relative to the photosphere; (d) flare frequency in each bin for all flares (red) and flares with larger equivalent duration than 1 (sec) (blue); and (e) flare energy in each bin for those. 
We discuss the result in comparison with other studies by photometry and spectroscopy in Section \ref{sec:photo}, correlation between spot locations and flare frequency in Section \ref{sec:phase}, two-year variation of the light curves for AU Mic and YZ CMi in Section \ref{sec:str}, and model selection and the validity for starspot modeling in Section \ref{sec:ml}.

\subsection{
Deduced parameters in comparison with other studies by photometry and spectroscopy} \label{sec:photo}
\begin{description}
\item[AU Mic ]\mbox{}\\ 
{In the two-spot models (Table \ref{tb:para1f} and \ref{tb:para1s}), the separation of spot longitude extends from} {$129.54 \pm 0.04$} {(deg) for Cycle 1 to} {$138.93 \pm 0.05$} {(deg) for Cycle 3. This slight change in the separation can be attributed to the differential rotation if assuming that the same spots have existed for two years and the star has significant differential rotation. In contrast, the deduced degree of differential rotation $\kappa$ for Cycle 1 and 3 are approximately 0, respectively.} 
This value of the differential rotation can not cause the change in the separation of spots. 
This may mean that the change in the relative longitude of two spots can be caused for different reasons rather than the differential rotation, although we need to pay attention to
this small value because the observation period of a TESS Sector ($\sim 27$ day) is not enough to deduce the differential rotation accurately.

{\cite{2019ApJ...883L...8W} also conducted starspot modeling for the same light curve with two spots using \texttt{STSP} code \citep[][]{2015ApJ...806..212D}. Their result of deduced spot latitudes $\simeq$ $9.6$ and $44.8$ (deg) are different from our results} {$ = -13.5^{+0.1}_{-0.2}$ and $39.7^{+0.1}_{-0.1}$} {(deg) probably due to the following three reasons:} (i) the high inclination angle $i = 75$ (deg) can result in large uncertainty in deducing spot latitudes,
 (ii) we sets the spot relative intensity $f_{\rm spot} = 0.43$ (Table \ref{tb:para0}) by calculating the intensity in TESS band based on the spot temperature deduced from the formula (Equation \ref{eq:spottemp} and \ref{eq:relspot}), whereas \cite{2019ApJ...883L...8W} assumes 0.7 \citep{2015ApJ...806..212D}, which results in the difference due to degeneracies between the spot relative intensity and latitudes \citepalias[][]{Ikuta20}, and (iii)
it is also difficult to discern whether a spot is on northern or southern hemisphere as the inclination angle increases \citep{2013ApJS..205...17W}.
{In addition, the difference of deduced spot longitude in our result} {$=129.54 \pm 0.04$} {(deg) is consistent with their result that the spots are separated by 131 (deg).} This consistency in the longitude is simply understood since the local minima of the light curve only correspond to the separation of spot longitude in the case of the two-spot model.

The surface map has been investigated through Doppler imaging and Zeeman Doppler imaging between Cycle 1 and 3 (September to November 2019) \citep{Klein20} and after Cycle 3 (November 2020 to September 2021) \citep{2022MNRAS.512.5067K}. As a result of Doppler imaging, \cite{Klein20} reported there were two dark features from the equator and 40 (deg) in addition to bright one around the equator in opposite to the dark one in 2019. \cite{2022MNRAS.512.5067K} also reported there were also dark features on the equator and 60 (deg) in 2020. {This results are different from our results of deduced latitudes for Cycle 1 and 3 in the two-spot model.} %$\Phi = -13.54 ^{+ 0.12 }_{- 0.11 }$ and $39.67 ^{+ 0.08 }_{- 0.07 }$ (deg) for Cycle 1 and $-0.55 ^{+ 0.18 }_{- 0.15 }$ and $80.22 ^{+ 0.01 }_{- 0.01 }$ (deg) for Cycle 3 in the two-spot model. 
It should be noted that the result of \cite{2019ApJ...883L...8W} is almost consistent with the results of Doppler imaging. One reason for this difference between our result and their result by Doppler imaging could be that there are degeneracies between the inclination angle and latitude, especially the high inclination angle $i = 75$ (deg) \citep{2013ApJS..205...17W}, in the starspot modeling.
We need to be careful to interpret these results obtained by both starspot modeling and Doppler imaging because these results largely depend on the mapping method and assumptions, such as prior distribution \citep{2017ApJ...849..120R,2021arXiv211006271L}.

{In the three-spot models (Table \ref{tb:para1f} and \ref{tb:para1s}), the latitudes in each model are almost the same absolute values.} Rotation period of each spot is almost the same value because of the almost the same latitude (Equation \ref{eq:rot}), and therefore the degree of differential rotation is not able to be measured accurately, {although large absolute values are deduced for Cycle 1 and 3.}

\item[YZ CMi]\mbox{}\\
{
In the two-spot models (Table \ref{tb:para2f} and \ref{tb:para2s}), the separation of spot longitude extends from} {$96.32 \pm 0.28$} {(deg) for Cycle 1 to} {$193.95 \pm 0.53$} {(deg) for Cycle 3. As the same as AU Mic, this slight change in the separation can be attributed to the differential rotation if assuming that the same spots have existed for two years and the star has significant differential rotation. In contrast, the deduced degree of differential rotation $\kappa$ are approximately 0, respectively. }
This may mean that the change in the relative longitude of two spots can be caused for different reasons rather than the differential rotation, although we need to pay attention to
this small value because the observation period of a TESS Sector ($\sim 27$ day) is not enough to deduce the differential rotation accurately, as discussed for AU Mic.

\cite{2022arXiv220611611B} also conducted starspot modeling for the same light curves with three and four spots for Cycle 1 and 3 using \texttt{BASSMAN} code, respectively. Their model is different in the point that they assume the inclination angle $=60$ (deg) and deduce the spot temperature (i.e, relative intensity) for each spot, although we set the angle $i=36$ (deg) and the spot temperature $T_{\rm spot}=2806$ (K) (relative intensity $f_{\rm spot} = 0.66$) for all spots. We do not respectively adopt three- and four-spot models for Cycle 1 and 3, and our result of deduced spot parameter is very different from their result. The difference is due to the models and assumptions, such as the number of spots (see  Section \ref{sec:ml}).

The surface map has been investigated through Doppler imaging before Cycle 1 (September 2016 to May 2017) \citep{Baroch20}. {The result suggests there are spots on the latitude from 75 to 81 (deg), and this result is consistent with our result of deduced latitudes} {$= 79.0^{+0.01}_{-0.02}$} {(deg) for Cycle 1 in the one-spot model.} %This could be due to the low inclination angle $i = 36$ (deg) because spots can be always visible.

\item[EV Lac]\mbox{}\\
The surface map has been investigated through Doppler imaging before Cycle 2 (January 2016 to December 2017) \citep[][]{2022arXiv220300415J}. {The result suggests that there are spots near the north pole, and is different from our result of deduced latitudes in both two- and three-spot models (Table \ref{tb:para3f}).}
\end{description}

\subsection{Spot location vs flare frequency} \label{sec:phase}
We investigate the correlation between spot locations and flare frequency in each rotation phase of the light curve. In Figure \ref{fig:phase1f}, \ref{fig:phase1s}, \ref{fig:phase2f}, \ref{fig:phase2s}, and \ref{fig:phase3f}, for each flares in the phase-folded light curve, 
we assume a null hypothesis that the flares are evenly distributed across ten bins of the rotation phase and perform a chi-squared test with nine degrees of freedom.
For AU Mic Cycle 1 and 3, YZ CMi Cycle 1 and 3, EV Lac Cycle 2, we obtain a reduced chi-square $=$ 0.899 (0.824), 0.381 (0.533), 0.505 (0.376), 0.439 (0.444), and 0.825 (0.890), and $p$-value $=$ 0.525 (0.594), 0.945 (0.851), 0.872 (0.947), 0.915 (0.911), and, 0.593 (0.533), for all flares (flares with larger equivalent duration than 1 sec), respectively.
Thus, the hypothesis that flares are evenly distributed across ten bins is not rejected, and we can not conclude that there is a correlation between the rotation phase of the light curve and occurrence of flares.
This result is consistent with the result in previous studies for different stars \citep{2014ApJ...797..121H, 2018MNRAS.480.2153D,2020MNRAS.494.3596D}.

In contrast, \cite{2022arXiv220611611B} suggested that the hypothesis that flares are evenly distributed in the rotation phase can not be rejected for 80 flares in Cycle 1 and can be rejected for 80 flares in Cycle 3. Our result is different from their result because more number of flares are detected for 124 in Cycle 1 and 137 in Cycle 3, respectively (Section \ref{sec:dataset}).

As a result of the starspot modeling, 
we can interpret the correlation between rotation phases and flare frequency as follows.
 Our result shows that either spot is always visible in all phases (Figure \ref{fig:phase1f}, \ref{fig:phase1s}, \ref{fig:phase2f}, \ref{fig:phase2s}, and \ref{fig:phase3f}) due to multiple spots or moderate inclination angle for each star. This suggests that flare frequency can be less correlated with the rotation phase if assuming that flares occur around spot visible to the line of sight at that time. Therefore, a polar spot is not necessarily needed to explain that the flares distribute in all rotation phases, and no or less correlation between rotation phases and flare frequency do not mean no relation between spot locations and flare occurrence.

\subsection{Two-year variation of light curves} \label{sec:str}
AU Mic and YZ CMi were observed in TESS Cycle 1 and 3 with the interval of approximately two years. 
It is shown that the amplitude of both light curves decreases (Figure \ref{fig:lc}), and the flare frequency is almost constant within an error (Figure \ref{fig:freq1} and \ref{fig:freq2})\footnote{We note that, when the flare frequency is compared, the flare frequency is need be adopted rather than the cumulative one because the cumulative one highly depends on large flares.}
In particular, the light curve of YZ CMi changes from one to two local minima per one equatorial rotation period.
Therefore, we compare the starspot modeling of Cycle 1 with that of Cycle 3 for AU Mic and YZ CMi in terms of spot latitude and size to explain the variations of light curves ascribed to spots in two years between different TESS Cycles.
{Then, we focus on the result of the two-spot model because two-spot models are adopted for Cycle 1 and 3.}

{
Comparing each of the two-spot models for AU Mic (Table \ref{tb:para1f} and \ref{tb:para1s}), deduced latitudes and radii are quite different. 
Although the latitude and radius of the first spot near the equator respectively correspond within an error of 14 and 3 (deg), the latitude of the second spot changes from mid-latitude to the pole.
The radii of the second spot on the pole is not determined due to the unspotted level of the light curve. We describe the unspotted level on the starspot modeling below.} 

{Comparing each of the two-spot models for YZ CMi (Table \ref{tb:para2f} and \ref{tb:para2s}), 
deduced latitudes and radii are slightly different.
The latitude of the first spot near the equator is different within an error of 25 (deg), and that of the second one at mid-latitude within an error of 5 (deg) because the latitude is determined by the visible duration of the spot \citep[][]{2013ApJS..205...17W}.
}
It is possible that the second spot decayed and the relative longitude changed due to the stellar differential rotation or the spot in Cycle 1 decayed and a new spot emerged in the interval of Cycle 1 and 3.

The total visible area relative to the stellar hemisphere ($2 \pi R^2_{\rm star}$) (Figure \ref{fig:phase1f}, \ref{fig:phase1s}, \ref{fig:phase2f}, and \ref{fig:phase2s}) changed from 0.076 to 0.142 for AU Mic and from 0.063 to 0.035 for YZ CMi. We found that the flare frequency did not change from Cycle 1 to Cycle 3 for both AU Mic and YZ CMi (Figure \ref{fig:freq1} and \ref{fig:freq2}). We suggest that this may be because the spot area deduced from the starspot modeling changes by a factor of two in two years, but this could not affect to change the flare frequency within the error. 

We note that it is difficult to determine the area of always visible spot on high latitude and the existence of a polar spot because such spot reduce the average flux and does not significantly contributes the modulation of the light curve. Thus, the unspotted level is not well determined in our result. The unspotted level of flux is of importance to estimate spot area and other spot parameters in the starspot modeling.
For each of the two-spot (three- or one-spot) models (Table \ref{tb:para1f}, \ref{tb:para1s}, \ref{tb:para2f}, \ref{tb:para2s}, and \ref{tb:para3f}), the reproduced average flux $F_{\rm ave}$ in the model are 0.831 (0.830), 0.816 (0.823), 0.839 (0.778), 0.847 (0.848), and 0.850 (0.842), respectively.
Then, the fluxes of Cycle 3 relative to those of 1 for AU Mic and YZ CMi are 0.982 and 1.010, respectively.
This reflects that the spot area has become smaller or a always visible polar spot that does not contribute to the brightness change has appeared, and  the amplitude of the light curve has become smaller from Cycle 1 to 3.
The reason that the radius of the polar spots reach the upper limit of the parameter range is that the unspotted level is not determined, and the same light curve can be reproduced by reducing the average flux. Therefore, it can be validated that there is only a polar spot in the two- and three-spot models for AU Mic Cycle 3. The possible large spot on the high latitude could affect the above conclusion (i.e., the change of the total spot area in two years), and more careful analysis may be necessary.

\subsection{Model selection and the validity for starspot modeling} \label{sec:ml}
More spots are indicated to be on the surface than the number of local minimum per one equatorial rotation period by analyses of spot occultations \citep[][]{2017ApJ...846...99M,2020ApJ...891..103N}.
In other words, the light curve of spotted stars always have two or one local minimum per one equatorial rotation period even if produced with any number of spots, but not two or one spot is actually present on the surface \citep[][]{2018ApJ...863..190B,2018ApJ...865..142B, 2021AJ....162..123L, 2021AJ....162..124L}.
Then, as well as \citetalias{Ikuta20}, we compare the number of spots based on the model selection by comparing the model evidence \citep{Kass1995}.
The values of the {logarithm of} model evidence 
$\log {\cal Z}$ for AU Mic Cycle 1 and 3, YZ CMi Cycle 1 and 3, and EV Lac Cycle 2 are listed in Table \ref{tb:para1f}, \ref{tb:para1s}, \ref{tb:para2f}, \ref{tb:para2s}, and \ref{tb:para3f}, respectively.

In cases of AU Mic Cycle 1 and 3, YZ CMi Cycle 3, and EV Lac Cycle 2, the three-spot model is much more decisive than the two-spot model by orders of magnitude: the evidences of the three-spot model relative to that of the two-spot model are $\Delta \log {\cal Z}=22568.926$, $53509.442$, $380.686$, and $3162.217$, respectively.
In the case of YZ CMi Cycle 1, the two-spot model is much more decisive than the one-spot model by orders of magnitude: the evidence of the two-spot model relative to that of the one-spot model is $\Delta \log {\cal Z}=10356.959$. In all cases, it is suggested that each of the model with one more additional spot is preferable in the Bayesian framework.
While we can also conduct starspot modeling with an additional spot, models with less than three spots are sufficient to investigate the correlation between the spot location and flare frequency (Section \ref{sec:phase}).
We note that the model evidence is affected by the choice of prior distributions \citep{BDA}. To constrain the number of spots precisely, it is also essential to scan the surface with spot occultations by multiple transits of exoplanets \citep{2017ApJ...846...99M, 2020ApJ...891..103N, 2021arXiv210406173B}.  

The peak-to-trough amplitude of light curves for AU Mic 1 and 3, YZ CMi 1 and 3, and EV Lac Cycle 2 are 0.046 (0.049), 0.032 (0.035), 0.035 (0.036), 0.014 (0.015), and 0.027 (0.028), for each of the two-spot model (three- or one-spot models), respectively. According to Figure 4 in \cite{2018ApJ...853..122R}, the peak-to-trough amplitude can be only explained by giant spots on the surface when the filling factors are approximately 0.22, 0.10, 0.12, 0.02, and 0.08, respectively. 
Since large spot groups are necessary to generate the large amplitude of light curves, it is sufficient to assume a few large spots for M-dwarf stars.

In addition, \cite{2017ApJ...849..120R} showed that the comparison between light curve inversion and Doppler imaging for a RS CVn star $\sigma$ Geminorum reveals qualitatively different surface map is obtained due to the various degeneracies inherent in each method.
Therefore, as future studies, multi-color and multi-spectrum mappings of the stellar surface are important to constrain the surface map more robustly.

\section{conclusion and future prospects} \label{sec:summary}
We conducted starspot modeling for TESS light curves of three M-dwarf flare stars, AU Mic, YZ CMi, and EV Lac, using the code implemented in \citetalias{Ikuta20}, for the purpose of investigating the spot properties and their relation with flare occurrences. 
The code enables to deduce multiple stellar/spot parameters by the adaptive parallel tempering algorithm efficiently.
The light curves of AU Mic Cycle 1 and 3, YZ CMi Cycle 3, and EV Lac Cycle 2 are optimized by the two- and three-spot models, and that of YZ CMi Cycle 1 by the two- and one-spot models. %(Figure \ref{fig:plot1f}, \ref{fig:plot1s}, \ref{fig:plot2f}, \ref{fig:plot2s}, and \ref{fig:plot3f}). 
For each model, unimodal posterior distributions are deduced (Table \ref{tb:para1f}, \ref{tb:para1s}, \ref{tb:para2f}, \ref{tb:para2s}, and \ref{tb:para3f}) regardless of the number of spots, and joint ones are delineated in Appendix \ref{sec:appendix}.
Flares are also detected (Figure \ref{fig:lc}) and accumulated as the flare frequency (Figure \ref{fig:freq1}, \ref{fig:freq2}, and \ref{fig:freq3}).
We discuss the results in terms of comparing with previous studies by photometry and spectroscopy, the relation between spot locations and flare frequency, the variation of the amplitude and shape of the light curve structures for AU Mic and YZ CMi in two years, the model selection for each numbers of model, and the validity of starspot modeling.

In Section \ref{sec:result}, we described the main results summarized as follows:

\begin{itemize}
    \item[(i)] The flare frequency is not necessarily correlated with rotation phase of the light curve for each star. We suggest that this can be because either spot is always visible in all phases for any model due to multiple spots or moderate inclination angle for each star. As the result, flares can be distributed in all rotation phases.
    \item[(ii)] For AU Mic and YZ CMi, the amplitude and shape of the light curve have varied in two years, whereas the flare frequency is almost constant within an error. The result of starspot modeling suggests that the variation of the light curve can be explained by the variations of spot size and latitude. We also found that 
    the total spot area does not change significantly in two years enough to change the flare frequency within the error, and this is why the flare frequency did not vary in two years.
\end{itemize}

In particular, EV Lac is to be observed in Sector 56 and 57 (September 1 to October 29 in 2022) through TESS second extended mission (Cycle 5). Then, three-year variation of the light curve can be investigated as in Section \ref{sec:str}.
In future studies, it is necessary to conduct surface mapping in multi-band wavelengths \citep[e.g.,][]{2018ApJ...857...39M,2020PASJ...72...68N, 2020ApJ...902...36T}, taking into account the effect of bright spots (faculae) and constraint on the spot temperature \citep[e.g.,][]{2021MNRAS.504.4751J}.
It is also helpful to scan the stellar surface with spot occultations by multiple transits of exoplanets \citep[e.g.,][]{2017ApJ...846...99M, 2020ApJ...891..103N, 2021arXiv210406173B}. 
Surface mapping on the stellar surface plays one of important roles in quantifying the
stellar effects on exoplanet characterizations \citep[][]{2018ApJ...853..122R,2022arXiv220109905R}.

\begin{acknowledgments}
{We sincerely appreciate the referee for providing feedback that helped to improve both content of this
manuscript and the clarity.} This study is based on public available data obtained by TESS mission. Funding for the
TESS mission is provided by NASA Science Mission Directorate. 
K.I. is grateful to Taichi Kato and Naoto Kojiguchi for their instructions of data analysis.
Numerical computations were carried out on Cray
XC40, Yukawa-21 (the Yukawa Institute for Theoretical Physics, Kyoto University), and Cray XC50 (Center for Computational Astrophysics, National Astronomical Observatory of Japan).
Our study was also supported by JSPS KAKENHI Grant Numbers: JP21H01131(all authors), JP18J20048, JP21J00316(K.N.), JP21J00106(Y.N.),
JP20K04032, JP20H05643, and JP20H00173(H.M.). 
{K.I. is supported by JST CREST Grant Number JPMJCR1761.}
Y.N. was supported by JSPS Overseas Research Fellowship Program {and NASA ADAP award program number 80NSSC21K0632 (PI: Adam Kowalski).} We also acknowledges the International Space Science Institute and the supported International Team 510: "Solar Extreme Events: Setting Up a Paradigm".
\end{acknowledgments}

\software{\texttt{stella} \citep{2020JOSS....5.2347F}, \texttt{corner} \citep{2016JOSS....1...24F}}

\appendix
\section{{Another models with different assumptions}} \label{sec:appendixb}
{
To evaluate the uncertainties of parameters (Section \ref{sec:set}), we also perform starspot modeling with the two-spot model for the light curve of EV Lac Cycle 2 with different assumptions: the quadratic limb-darkening law in Section \ref{sec:a}, parameters including the spot relative intensity under uniform and normal prior distributions in Section \ref{sec:ab}, and the two-spot model with systematic noise model by Gaussian process in Section \ref{sec:a3}. The deduced parameters are listed in Table \ref{tb:para3s}.

\subsection{Quadratic limb-darkening law} \label{sec:a}
To compare the result by the quadratic limb-darkening law with that by the nonlinear law (Table \ref{tb:para0}), we also fix the coefficient values $c_1=c_3=0$, $c_2=u_1+2u_2=1.04$, and $c_4=-u_2=-0.44$ in Equation \ref{eq:limb}, where $u_1=0.16$ and $u_2=0.44$ (Equation 17 in \citetalias{Ikuta20}) from \cite{Claret18}.
%The uncertainties of the coefficients are not calculated in \cite{Claret18}, and we do not include the coefficients in the parameters but fix the values.
As a result, 
the uncertainties of spot parameters are the same degree with those in the two-spot model (Table \ref{tb:para3f}).
The uncertainties of the stellar parameters increase by an order of magnitude, but remain small.
The deduced latitude of the second spot is lower because there is degeneracies between the spot latitude and size due to the high inclination angle $=60$ (deg) (\citetalias{Ikuta20}).

\subsection{Spot relative intensity with uniform and normal prior distributions} \label{sec:ab}
To take account of the degeneracies between spot latitude and spot relative intensity, we also include it as parameters under uniform and normal prior distributions extended to the range from 0 to 2 for bright spots as reported in \cite{2018ApJ...857...39M}.
As a result, the uncertainties of stellar parameters are also the same degree with those in the two-spot model (Table \ref{tb:para3f}).
The uncertainties of spot parameters increase by a factor, but remain small.
The deduced latitude of the second spot is lower for the same reason in Section \ref{sec:a}, and the deduced radii are larger because the deduced relative intensity is larger (Section \ref{sec:set}).
The deduced spot intensities $f_{\rm spot}=0.833$ and $0.811$ for the uniform and normal prior distributions correspond to  
the spot temperature $T_{\rm spot}= 3284$ and $3268$ (K), respectively (Equation \ref{eq:relspot}).

\subsection{Two-spot and systematic noise models} \label{sec:a3}
To evaluate systematic noise in the residual from the two-spot model, and we simultaneously perform modeling of the noise with Gaussian process \citep[GP;][]{2017AJ....154..220F,2021ApJS..254...13G}. 
Then, we employ a multivariate normal likelihood function 
with the Mat\'{e}rn-3/2 kernel characterized by hyperparameters $\sigma_{\rm sys}$ and $\rho_{\rm sys}$:
\begin{equation}
p ({\cal D}|\theta) = \frac{1}{(\sqrt{2\pi})^N \sqrt{|{\boldsymbol K}+\sigma_i^2 {\boldsymbol I}|}} \exp{\biggl[ 
-\frac{1}{2} \bigl({\boldsymbol F}_{{\rm obs}}-{\boldsymbol F}_{{\rm mod}}({\theta})\bigr)^{\rm T} ({\boldsymbol K}+\sigma_i^2 {\boldsymbol I})^{-1} \bigl({\boldsymbol F}_{{\rm obs}}-{\boldsymbol F}_{{\rm mod}}({\theta})\bigr)
\biggr]}
\end{equation}

and
\begin{equation}
K(t_i,t_j) = \sigma_{\rm sys}^2 \bigl( 1 + \frac{\sqrt{3} |t_i-t_j|}{\rho_{\rm sys}} \bigr) \exp \bigl( -\frac{\sqrt{3} |t_i-t_j|}{\rho_{\rm sys}} \bigr),
\end{equation}

where $\theta$, ${\cal D}$, $\sigma_i$, ${\boldsymbol F}_{{\rm obs}}$, and ${\boldsymbol F}_{{\rm mod}}$ denote parameters, observed data, photometric error, the vector of the observed flux, and  that of the model flux, respectively.
As a result, the uncertainties of spot parameters are slightly larger than those in the two-spot model (Table \ref{tb:para3f}) up to an order of magnitude because the systematic noise model enables to optimize the residual from the spot model flexibly.
The uncertainties of stellar parameters increase by two order of magnitude, but remain small.
The deduced latitude of the second spot is also lower for the same reason in Section \ref{sec:a}.

}

\begin{figure*}[p]  
%\epsscale{0.9}
\plotone{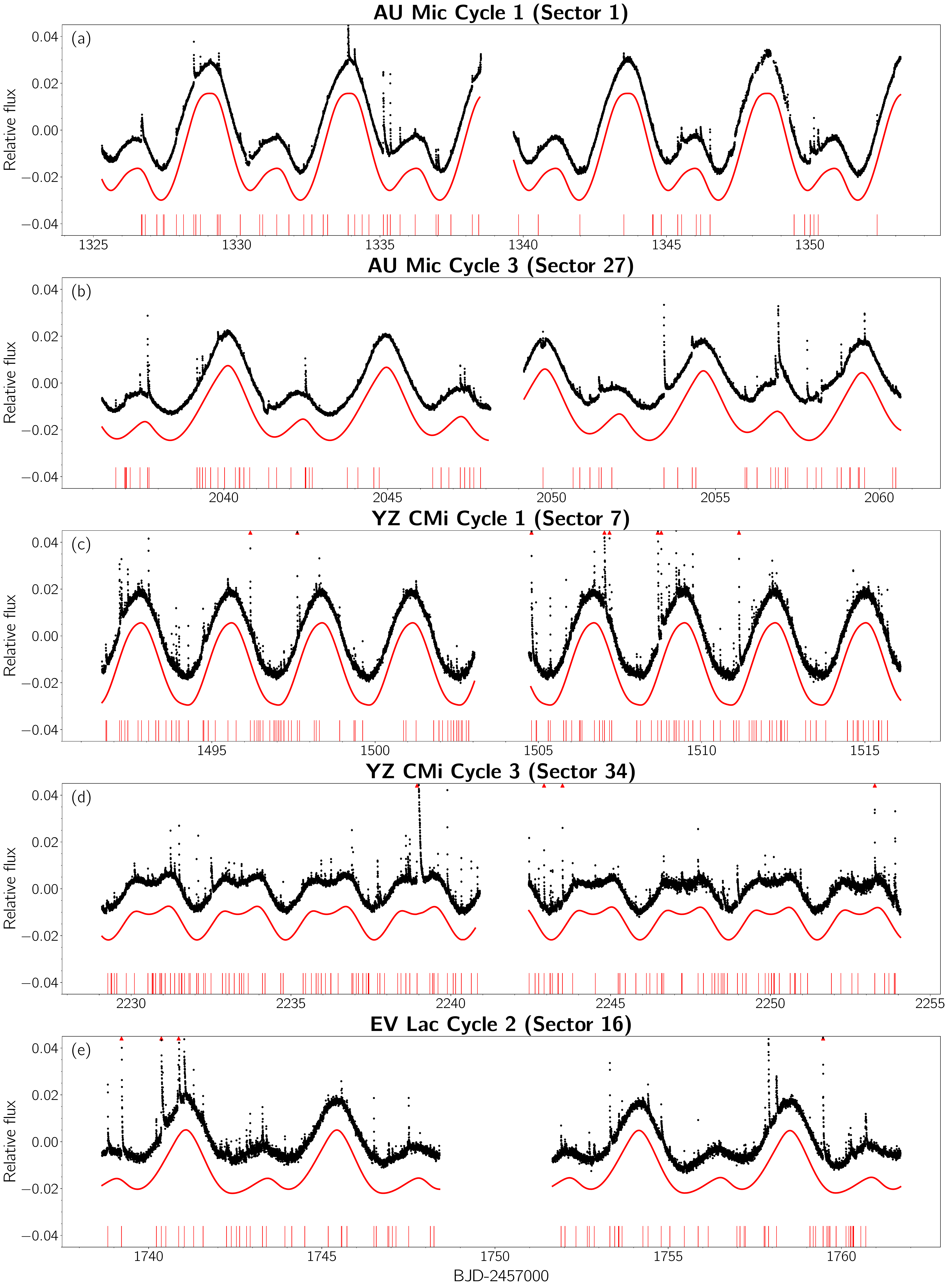} 
\caption{(a, b, c, d, e) TESS light curves of AU Mic Cycle 1 (Sector 1, July 25 to August 22 in 2018) and 3 (Sector 27, July 4 to 30 in 2020), YZ CMi Cycle 1 (Sector 7, January 7 to February 2 in 2019) and 3 (Sector 34, January 13 to February 9), and EV Lac Cycle 2 (Sector 16, September 11 to October 7 in 2019) (black), respectively. The relative flux is calculated by the PDC-SAP flux normalized by its average excluding missing values. {Detected flares are represented in each peak time (red bar) (for details, Section \ref{sec:dataset}), and the large flares are marked as outliers (red triangle). The reproduced light curves with the two-spot models are also exhibited under the TESS light curves (red).} \label{fig:lc}} 
\end{figure*}

\begin{figure*}[p]
\epsscale{0.5}
\plotone{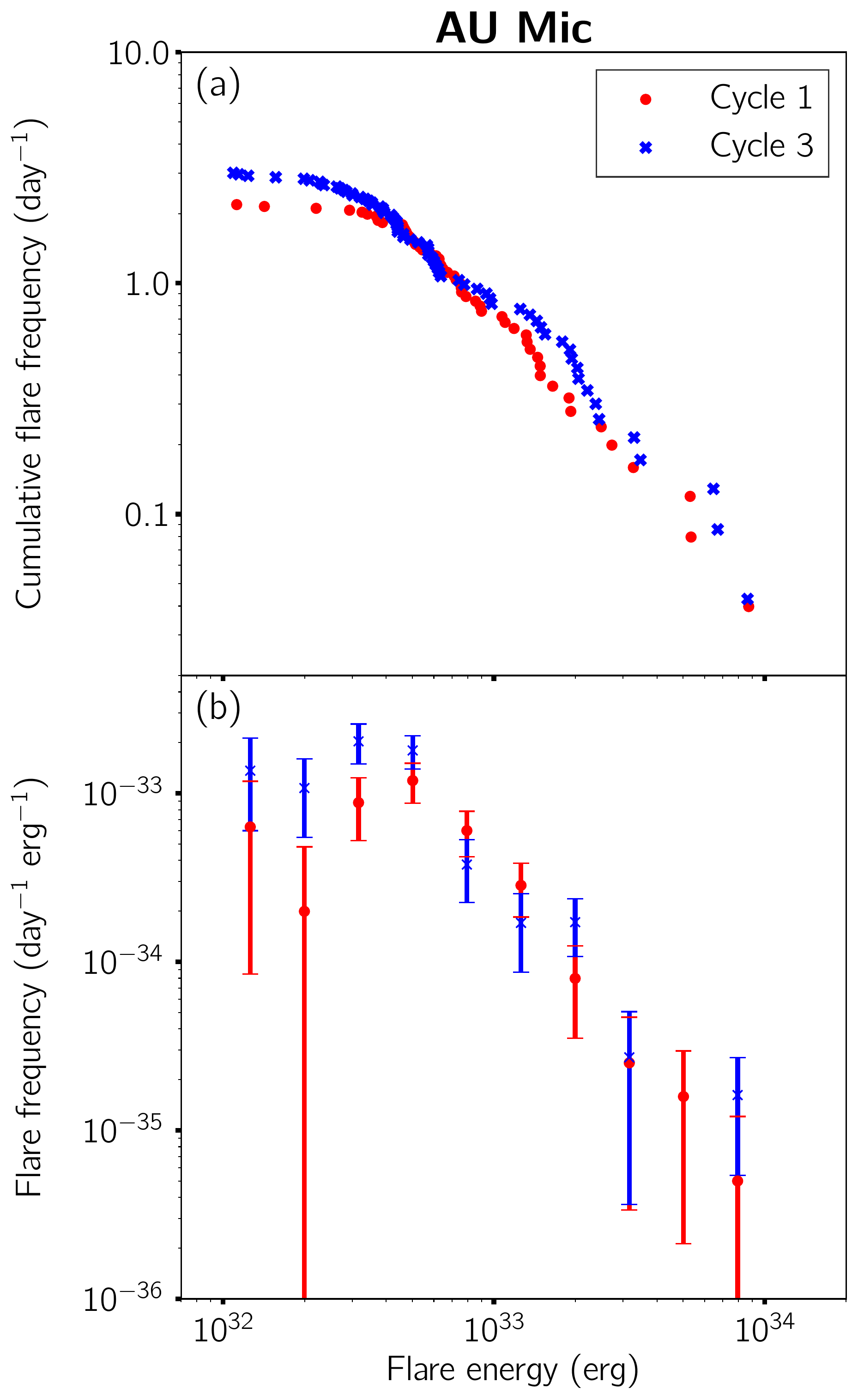} 
\caption{Flare frequency for AU Mic Cycle 1 (red) and 3 (blue). The number of flares per observation period of a TESS Sector (Sector 1 and 27) are 55/25.131 = 2.189 and 70/23.307 = 3.003 (day$^{-1}$), respectively. (a) Cumulative flare frequency distribution versus bolometric flare energy: the number of flares per day with an energy larger than the energy value. (b)
Flare frequency distribution versus bolometric flare
energy: the number of flares per day and energy in each energy bin. The error bar equals an square root of (the number of flares in each energy bin + 1) per observation period and energy in each energy bin. \label{fig:freq1}}
\end{figure*}

\begin{figure*}[p]
\epsscale{0.5}
\plotone{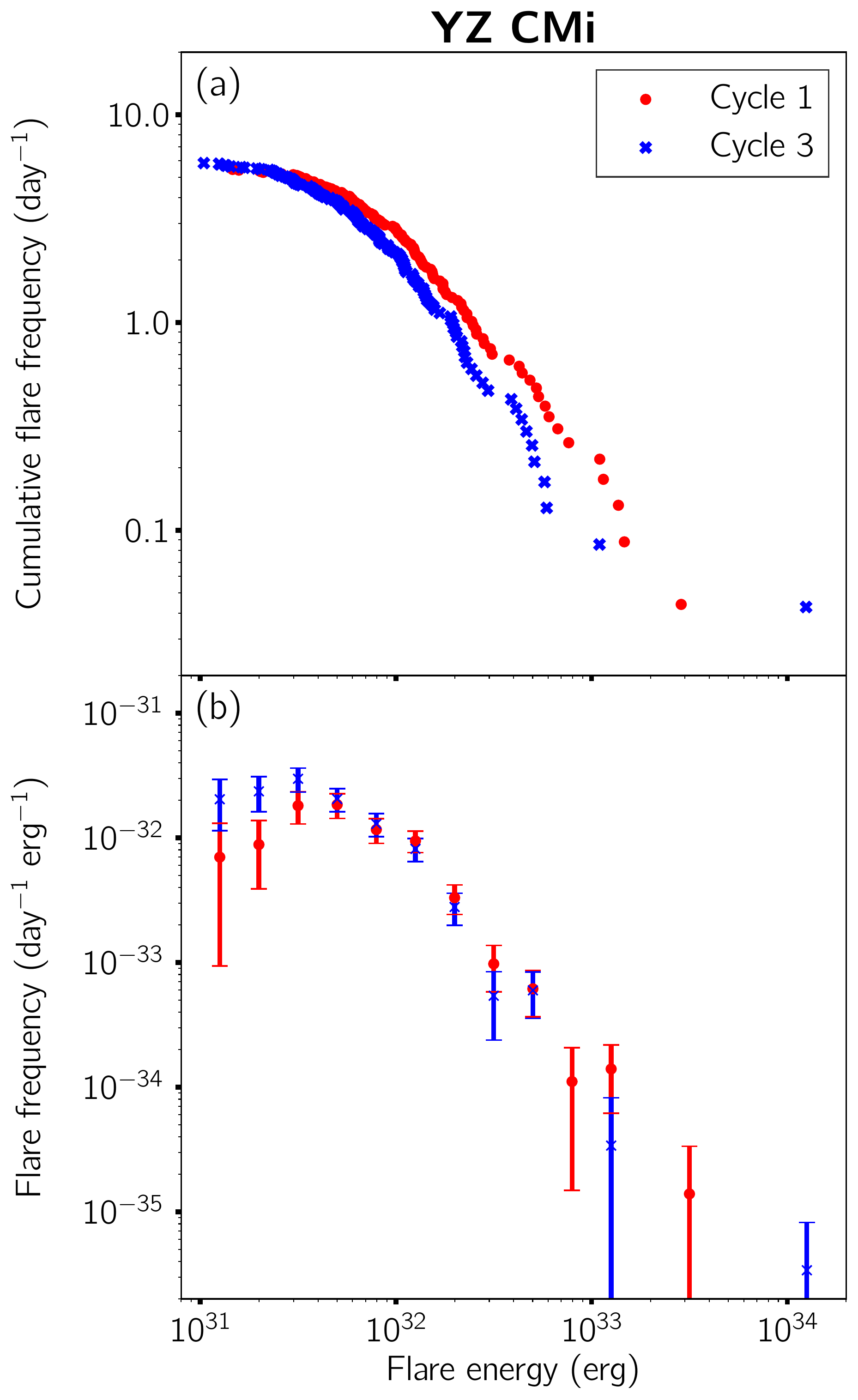} 
\caption{Same as Figure \ref{fig:freq1} but for YZ CMi Cycle 1 and 3. The number of flares per observation period of a TESS Sector (Sector 7 and 34) are 124/22.725 = 5.457 and 137/23.398 = 5.855 (day$^{-1}$), respectively.} \label{fig:freq2}
\end{figure*}

\begin{figure*}[p]
\epsscale{0.5}
\plotone{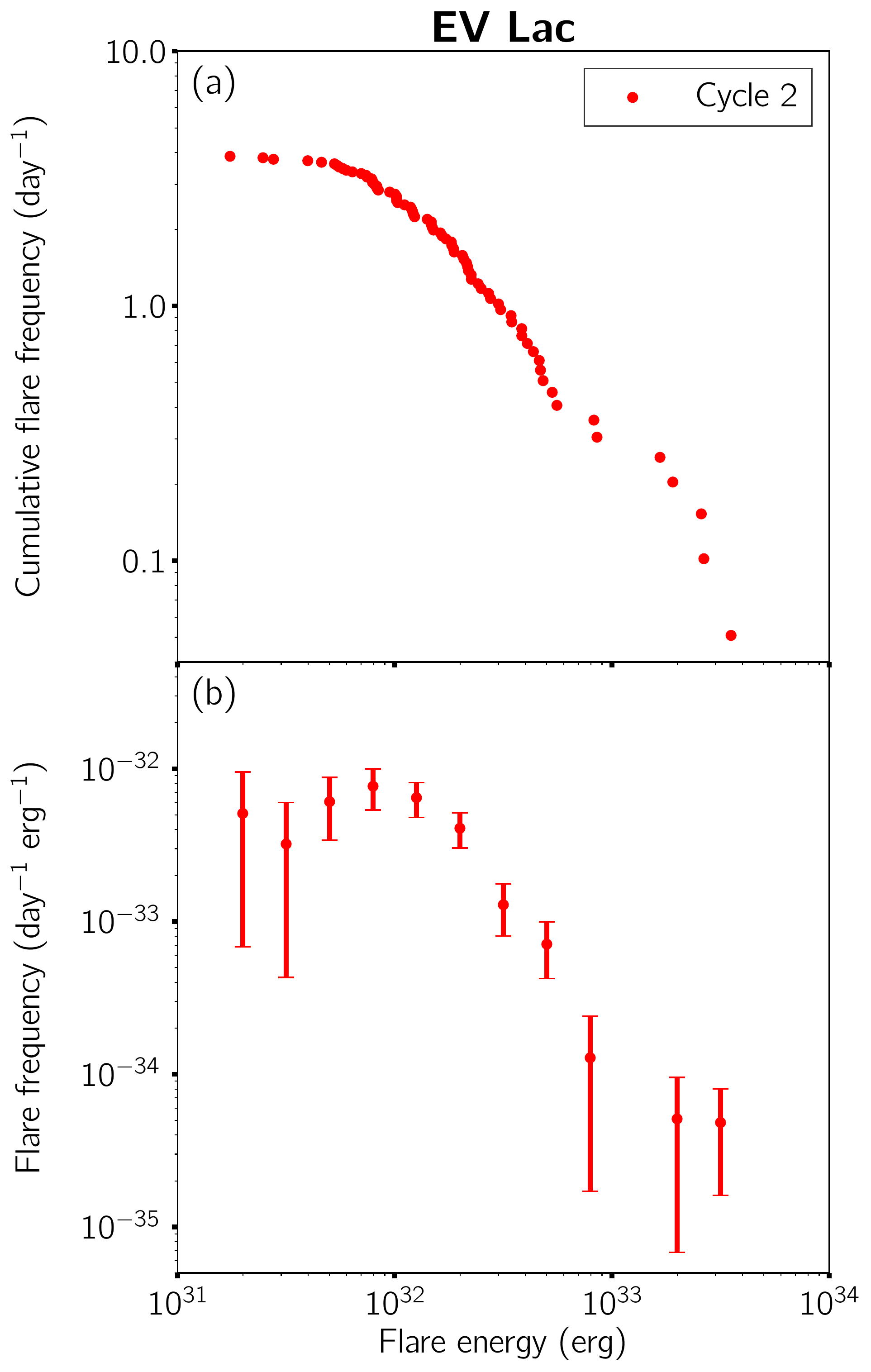} 
\caption{Same as Figure \ref{fig:freq1} but for EV Lac Cycle 2. The number of flares per observation period of a TESS Sector (Sector 16) is 76/19.650 = 3.868 (day$^{-1}$).}\label{fig:freq3} 
\end{figure*}

\begin{deluxetable*}{lccc}[p]
\tablecaption{AU Mic Cycle 1 case \label{tb:para1f}}
\tabletypesize{\footnotesize}
\tablehead{
\colhead{Deduced Parameters} &  \colhead{Two-spot Model} & \colhead{{\bf Three-spot Model}} & \colhead{Prior Distribution\tablenotemark{a}}
}
\startdata
(Stellar parameters) &&& \\
1.  Equatorial period $P_{{\scriptsize \textrm{eq}}}$ (day) &$4.8624 ^{+ 0.0001 }_{- 0.0001 }$&$4.8089 ^{+ 0.0010 }_{- 0.0006 }$&${\cal U}_{\log} (4.5000,5.0000)$  \\
2.   Degree of differential rotation $\kappa$ & 
$-0.0001 ^{+ 0.0001 }_{- 0.0001 }$&$0.1998 ^{+ 0.0002 }_{- 0.0022 }$&${\cal U} (-0.2000,0.2000)$  \\
(Spot parameters) &  & &   \\
(1st spot) &  &  &  \\
3.     Latitude $\Phi_1$  (deg) &
$-13.54 ^{+ 0.12 }_{- 0.11 }$&$-14.72 ^{+ 0.11 }_{- 0.07 }$&{${\cal U} (-90.00,\Phi_2)$\tablenotemark{b}} \\
4.     Initial longitude $\Lambda_1$  (deg)&$-22.95 ^{+ 0.02 }_{- 0.03 }$&$-134.06 ^{+ 0.05 }_{- 0.06 }$&${\cal U}  (-180.00,180.00)$ \\
5.     Maximum radius $\alpha_{{\scriptsize \textrm{max,1}}}$  (deg)&$15.61 ^{+ 0.01 }_{- 0.02 }$&$12.86 ^{+ 0.02 }_{- 0.01 }$ &${\cal U} (0.01,30.00)$  \\
(2nd spot)&  &  &   \\
6.     Latitude $\Phi_2$  (deg) &$39.67 ^{+ 0.08 }_{- 0.07 }$&$12.08 ^{+ 0.13 }_{- 0.09 }$&{${\cal U} (\Phi_1,\Phi_3)$\tablenotemark{b}} \\
7.     Initial longitude $\Lambda_2$ (deg) & $-152.49 ^{+ 0.02 }_{- 0.02 }$&$179.39 ^{+ 0.06 }_{- 0.12 }$&${\cal U}  (-180.00,180.00)$\\
8.     Maximum radius $\alpha_{{\scriptsize \textrm{max,2}}}$ (deg)&$16.02 ^{+ 0.01 }_{- 0.01}$&$10.90 ^{+ 0.01 }_{- 0.02 }$&${\cal U}(0.01,30.00)$\\
(3rd spot) &  &  & \\
9.     Latitude $\Phi_3$  (deg) &-& $13.92 ^{+ 0.10 }_{- 0.10 }$ &${\cal U} (\Phi_2,90.00)$\tablenotemark{b} \\
10.     Initial longitude $\Lambda_3$ (deg)  &-&$-24.31 ^{+ 0.03 }_{- 0.02 }$&${\cal U}  (-180.00,180.00)$  \\
11.     Maximum radius $\alpha_{{\scriptsize \textrm{max,3}}}$ (deg) &-& $14.47 ^{+ 0.01}_{- 0.01}$&${\cal U}(0.01,30.00)$ \\ \hline
Average flux $F_{\rm ave}$ & 0.831 & 0.830 & \\ \hline
{Logarithm of model} evidence $\log{\cal Z}$ & $-63950.957$& {$\boldsymbol{-41382.031}$}&   \\  
\enddata
\tablenotetext{a}{
{${\cal U_{\text{log}}}(a,b)= 1/(\theta \log(b/a))$} and ${\cal U}(a,b)=1/(b-a)$ represent bounded log uniform distribution (Jeffreys prior) and bounded uniform distribution defined in $a\leq \theta \leq b$, respectively.
}
\tablenotetext{b}{
We discern each spot by its latitude $\Phi_k$, not by its reference time $t_k$ in \citetalias{Ikuta20}, to improve the sampling efficiency of the PT. In case of the two-spot model, we set $\Phi_3 = 90.0$ (the upper limit of the latitude).
}
\end{deluxetable*}

\begin{deluxetable*}{lccc}[p]
\tablecaption{AU Mic Cycle 3 case \label{tb:para1s}}
\tabletypesize{\footnotesize}
\tablehead{
\colhead{Deduced Parameters} &  \colhead{Two-spot Model} & \colhead{{\bf Three-spot Model}} & \colhead{Prior Distribution\tablenotemark{a}}  
}
\startdata
(Stellar parameters) &&& \\
1.  Equatorial period $P_{{\scriptsize \textrm{eq}}}$ (day) &$4.8289 ^{+ 0.0001 }_{- 0.0002 }$&$5.2004 ^{+ 0.0112 }_{- 0.0061 }$&${\cal U}_{\log} (4.5000,5.0000)$\\
2.   Degree of differential rotation $\kappa$ & $0.0094 ^{+ 0.0001 }_{- 0.0001 }$&$-0.1072 ^{+ 0.0018 }_{- 0.0032 }$&${\cal U} (-0.2000,0.2000)$\\
(Spot parameters) &  & &   \\
(1st spot) &  &  &  \\
3.     Latitude $\Phi_1$  (deg) &$-0.55 ^{+ 0.18 }_{- 0.15 }$&$-59.04 ^{+ 0.06 }_{- 0.06 }$&{${\cal U} (-90.00,\Phi_2)$\tablenotemark{b}}\\
4.     Initial longitude $\Lambda_1$  (deg)&$-14.36 ^{+ 0.03 }_{- 0.02 }$&$-1.32 ^{+ 0.06 }_{- 0.07 }$&${\cal U}  (-180.00,180.00)$\\
5.     Maximum radius $\alpha_{{\scriptsize \textrm{max,1}}}$  (deg)&$12.80 ^{+ 0.01 }_{- 0.01 }$&$30.00 ^{+ 0.00 }_{- 0.01}$ &${\cal U} (0.01,30.00)$\\
(2nd spot)&  &  &   \\
6.     Latitude $\Phi_2$  (deg) &$80.22 ^{+ 0.01 }_{- 0.01 }$&$52.57 ^{+ 0.12 }_{- 0.10 }$&{${\cal U} (\Phi_1,\Phi_3)$\tablenotemark{b}}\\
7.     Initial longitude $\Lambda_2$ (deg) & $-153.29 ^{+ 0.04 }_{- 0.03 }$&$164.28 ^{+ 0.13 }_{- 0.14 }$&${\cal U}  (-180.00,180.00)$\\
8.     Maximum radius $\alpha_{{\scriptsize \textrm{max,2}}}$ (deg)&$30.00 ^{+ 0.00 }_{- 0.01 }$&$16.42 ^{+ 0.02 }_{- 0.04 }$&${\cal U}(0.01,30.00)$\\
(3rd spot) &  &  & \\
9.     Latitude $\Phi_3$  (deg) &-& $56.28 ^{+ 0.05 }_{- 0.05 }$ &${\cal U} (\Phi_2,90.00)$\tablenotemark{b}\\
10.     Initial longitude $\Lambda_3$ (deg)  &-&$-96.59 ^{+ 0.09 }_{- 0.11 }$&${\cal U}  (-180.00,180.00)$\\
11.     Maximum radius $\alpha_{{\scriptsize \textrm{max,3}}}$ (deg) &-& $16.13 ^{+ 0.03 }_{- 0.03 }$&${\cal U}(0.01,30.00)$\\ \hline
Average flux $F_{\rm ave}$ & 0.816 & 0.823  & \\ \hline
{Logarithm of model} evidence $\log{\cal Z}$ & $-139256.404$& {$\boldsymbol{-85746.962}$} & \\
\enddata
\tablenotetext{a}{
{${\cal U_{\text{log}}}(a,b)= 1/(\theta \log(b/a))$} and ${\cal U}(a,b)=1/(b-a)$ represent bounded log uniform distribution (Jeffreys prior) and bounded uniform distribution defined in $a\leq \theta \leq b$, respectively.
}
\tablenotetext{b}{
We discern each spot by its latitude $\Phi_k$, not by its reference time $t_k$ in \citetalias{Ikuta20}, to improve the sampling efficiency of the PT. In case of the two-spot model, we set $\Phi_3 = 90.0$ (the upper limit of the latitude).
}
\end{deluxetable*}

\begin{deluxetable*}{lccc}[p]
\tablecaption{YZ CMi Cycle 1 case \label{tb:para2f}} 
\tabletypesize{\footnotesize}
\tablehead{
\colhead{Deduced Parameters} &  \colhead{{\bf Two-spot Model}} &  \colhead{One-spot Model} & \colhead{Prior Distribution\tablenotemark{a}}  
}
\startdata
(Stellar parameters) &&& \\
1.  Equatorial period $P_{{\scriptsize \textrm{eq}}}$ (day) &$2.7753 ^{+ 0.0002 }_{- 0.0003 }$&$2.7737 ^{+ 0.0001 }_{- 0.0001 }$&${\cal U}_{\log} (2.0000,3.0000)$\\
2.   Degree of differential rotation $\kappa$ & $-0.0008 ^{+ 0.0002 }_{- 0.0002 }$&$0.0000$ (\textit{fixed}) &${\cal U} (-0.2000,0.2000)$\\
(Spot parameters) &  & &   \\
(1st spot) &  &  &  \\
3.     Latitude $\Phi_1$  (deg) &$-6.60 ^{+ 0.24 }_{- 0.19 }$&$79.03 ^{+ 0.01 }_{- 0.02 }$&{${\cal U} (-90.00,\Phi_2)$\tablenotemark{b}}\\
4.     Initial longitude $\Lambda_1$  (deg)&$113.68 ^{+ 0.20 }_{- 0.22 }$&$31.35 ^{+ 0.06 }_{- 0.06 }$&${\cal U}  (-180.00,180.00)$\\
5.     Maximum radius $\alpha_{{\scriptsize \textrm{max,1}}}$  (deg)&$12.73 ^{+ 0.07 }_{- 0.08 }$&$29.99 ^{+ 0.01 }_{- 0.02 }$&${\cal U} (0.01,30.00)$\\
(2nd spot)&  &  &   \\
6.     Latitude $\Phi_2$  (deg) &$41.17 ^{+ 0.30 }_{- 0.28 }$&-&{${\cal U} (\Phi_1,90.00)$\tablenotemark{b}}\\
7.     Initial longitude $\Lambda_2$ (deg) & 
$17.36 ^{+ 0.12 }_{- 0.18 }$& -&${\cal U}  (-180.00,180.00)$\\
8.     Maximum radius $\alpha_{{\scriptsize \textrm{max,2}}}$ (deg)&$15.92 ^{+ 0.01 }_{- 0.02 }$&-&${\cal U}(0.01,30.00)$\\ \hline
Average flux $F_{\rm ave}$ & 0.839 & 0.778  & \\ \hline
{Logarithm of model} evidence $\log{\cal Z}$ &{$\boldsymbol{14982.598}$}&$4625.639$& \\
\enddata
\tablenotetext{a}{
{${\cal U_{\text{log}}}(a,b)= 1/(\theta \log(b/a))$} and ${\cal U}(a,b)=1/(b-a)$ represent bounded log uniform distribution (Jeffreys prior) and bounded uniform distribution defined in $a\leq \theta \leq b$, respectively.
}
\tablenotetext{b}{
We discern each spot by its latitude $\Phi_k$, not by its reference time $t_k$ in \citetalias{Ikuta20}, to improve the sampling efficiency of the PT. In case of the one-spot model, we set $\Phi_2 = 90.0$ (the upper limit of the latitude).
}
\end{deluxetable*}

\begin{deluxetable*}{lccc}[p]
\tablecaption{YZ CMi Cycle 3 case \label{tb:para2s}}
\tabletypesize{\footnotesize}
\tablehead{
\colhead{Deduced Parameters} &  \colhead{Two-spot Model} & \colhead{{\bf Three-spot Model}} & \colhead{Prior Distribution\tablenotemark{a}}  
}
\startdata
(Stellar parameters) &&& \\
1.  Equatorial period $P_{{\scriptsize \textrm{eq}}}$ (day) &$2.7705 ^{+ 0.0003 }_{- 0.0002 }$&$2.7707 ^{+ 0.0003 }_{- 0.0003 }$
&${\cal U}_{\log} (2.0000,3.0000)$\\
2.   Degree of differential rotation $\kappa$ &$0.0082 ^{+ 0.0007 }_{- 0.0008 }$&$0.0088 ^{+ 0.0016 }_{- 0.0011 }$& ${\cal U} (-0.2000,0.2000)$\\
(Spot parameters) &  & &  \\
(1st spot) &  &  &  \\
3.     Latitude $\Phi_1$  (deg) &$18.48 ^{+ 0.28 }_{- 0.30 }$&$-10.92 ^{+ 0.50 }_{- 0.45 }$ &{${\cal U} (-90.00,\Phi_2)$\tablenotemark{b}}\\
4.     Initial longitude $\Lambda_1$  (deg)& $-23.68 ^{+ 0.17 }_{- 0.15 }$&$133.41 ^{+ 0.81 }_{- 0.66 }$&${\cal U}  (-180.00,180.00)$\\
5.     Maximum radius $\alpha_{{\scriptsize \textrm{max,1}}}$  (deg)&$12.91 ^{+ 0.06 }_{- 0.05 }$&$7.91 ^{+ 0.14 }_{- 0.19 }$&${\cal U} (0.01,30.00)$\\
(2nd spot)&  &  &   \\
6.     Latitude $\Phi_2$  (deg) &$38.82 ^{+ 1.21 }_{- 0.81 }$&$16.36 ^{+ 0.36 }_{- 0.29 }$&{${\cal U} (\Phi_1,\Phi_3)$\tablenotemark{b}}\\
7.     Initial longitude $\Lambda_2$ (deg) &$170.27 ^{+ 0.50 }_{- 0.50 }$&$-17.50 ^{+ 0.35 }_{- 0.52 }$&${\cal U}  (-180.00,180.00)$\\
8.     Maximum radius $\alpha_{{\scriptsize \textrm{max,2}}}$ (deg)&$7.77 ^{+ 0.12 }_{- 0.10 }$&$12.34 ^{+ 0.05 }_{- 0.05 }$&${\cal U}(0.01,30.00)$\\
(3rd spot) &  &  & \\
9.     Latitude $\Phi_3$  (deg) &-& $49.48 ^{+ 5.76 }_{- 2.69 }$ &${\cal U} (\Phi_2,90.00)$\tablenotemark{b}\\
10.     Initial longitude $\Lambda_3$ (deg)  &-& 
$-146.64 ^{+ 2.46 }_{- 2.57 }$ &${\cal U}  (-180.00,180.00)$\\
11.     Maximum radius $\alpha_{{\scriptsize \textrm{max,3}}}$ (deg) &-&$6.59 ^{+ 0.41 }_{- 0.18 }$ &${\cal U}(0.01,30.00)$\\ \hline
Average flux $F_{\rm ave}$ & 0.847 & 0.848 & \\ \hline
{Logarithm of model} evidence $\log{\cal Z}$ &$13111.858$ & {$\boldsymbol{13492.544}$} & \\
\enddata
\tablenotetext{a}{
{${\cal U_{\text{log}}}(a,b)= 1/(\theta \log(b/a))$} and ${\cal U}(a,b)=1/(b-a)$ represent bounded log uniform distribution (Jeffreys prior) and bounded uniform distribution defined in $a\leq \theta \leq b$, respectively.
}
\tablenotetext{b}{
We discern each spot by its latitude $\Phi_k$, not by its reference time $t_k$ in \citetalias{Ikuta20}, to improve the sampling efficiency of the PT. In case of the two-spot model, we set $\Phi_3 = 90.0$ (the upper limit of the latitude).
}
\end{deluxetable*}

\begin{deluxetable*}{lccc}[p]
\tablecaption{EV Lac Cycle 2 case \label{tb:para3f}}
\tabletypesize{\footnotesize}
\tablehead{
\colhead{Deduced Parameters} &  \colhead{Two-spot Model} & \colhead{{\bf Three-spot Model}} & \colhead{Prior Distribution\tablenotemark{a}}  
}
\startdata
(Stellar parameters) &&& \\
1.  Equatorial period $P_{{\scriptsize \textrm{eq}}}$ (day) &$4.3573 ^{+ 0.0002 }_{- 0.0003 }$&$4.3651 ^{+ 0.0003 }_{- 0.0002 }$&${\cal U}_{\log} (4.0000,5.0000 )$\\
2.   Degree of differential rotation $\kappa$ & $0.0020 ^{+ 0.0001 }_{- 0.0001 }$&$-0.0033 ^{+ 0.0001 }_{- 0.0001 }$&${\cal U} (-0.2000,0.2000)$\\
(Spot parameters) &  & &   \\
(1st spot) &  &  &  \\
3.     Latitude $\Phi_1$  (deg) &$24.89 ^{+ 0.28 }_{- 0.21 }$&$9.91 ^{+ 0.12 }_{- 0.18 }$&{${\cal U} (-90.00,\Phi_2)$\tablenotemark{b}}\\
4.     Initial longitude $\Lambda_1$  (deg)&$60.35 ^{+ 0.11 }_{- 0.13 }$&$-113.64 ^{+ 0.20 }_{- 0.20 }$&${\cal U}  (-180.00,180.00)$\\
5.     Maximum radius $\alpha_{{\scriptsize \textrm{max,1}}}$  (deg)&$13.44 ^{+ 0.02 }_{- 0.02 }$&$14.67 ^{+ 0.04 }_{- 0.06 }$&${\cal U} (0.01,30.00)$\\
(2nd spot)&  &  &   \\
6.     Latitude $\Phi_2$  (deg) &$55.89 ^{+ 0.31 }_{- 0.26 }$&$9.95 ^{+ 0.16 }_{- 0.13 }$&{${\cal U} (\Phi_1,\Phi_3)$\tablenotemark{b}}\\
7.     Initial longitude $\Lambda_2$ (deg) &$-84.11 ^{+ 0.11 }_{- 0.11 }$&$111.85 ^{+ 0.32 }_{- 0.35 }$&${\cal U}  (-180.00,180.00)$\\
8.     Maximum radius $\alpha_{{\scriptsize \textrm{max,2}}}$ (deg)&$14.36 ^{+ 0.06 }_{- 0.06 }$&$10.56 ^{+ 0.07 }_{- 0.09 }$&${\cal U}(0.01,30.00)$\\
(3rd spot) &  &  & \\
9.     Latitude $\Phi_3$  (deg) &-&$54.15 ^{+ 0.53 }_{- 0.47 }$&${\cal U} (\Phi_2,90.00)$\tablenotemark{b}\\
10.     Initial longitude $\Lambda_3$ (deg)  &-&$13.71 ^{+ 0.36 }_{- 0.25 }$&${\cal U}  (-180.00,180.00)$\\
11.     Maximum radius $\alpha_{{\scriptsize \textrm{max,3}}}$ (deg) &-&$15.99 ^{+ 0.06 }_{- 0.04 }$
&${\cal U}(0.01,30.00)$\\ \hline
Average flux $F_{\rm ave}$ & 0.850 & 0.842  & \\ \hline
{Logarithm of model} evidence $\log{\cal Z}$ & $-7297.802$& {$\boldsymbol{-4135.585}$} & \\
\enddata
\tablenotetext{a}{
{${\cal U_{\text{log}}}(a,b)= 1/(\theta \log(b/a))$} and ${\cal U}(a,b)=1/(b-a)$ represent bounded log uniform distribution (Jeffreys prior) and bounded uniform distribution defined in $a\leq \theta \leq b$, respectively.
}
\tablenotetext{b}{
We discern each spot by its latitude $\Phi_k$, not by its reference time $t_k$ in \citetalias{Ikuta20}, to improve the sampling efficiency of the PT. In case of the two-spot model, we set $\Phi_3 = 90.0$ (the upper limit of the latitude).
}
\end{deluxetable*}

\begin{figure*}[p]
\plotone{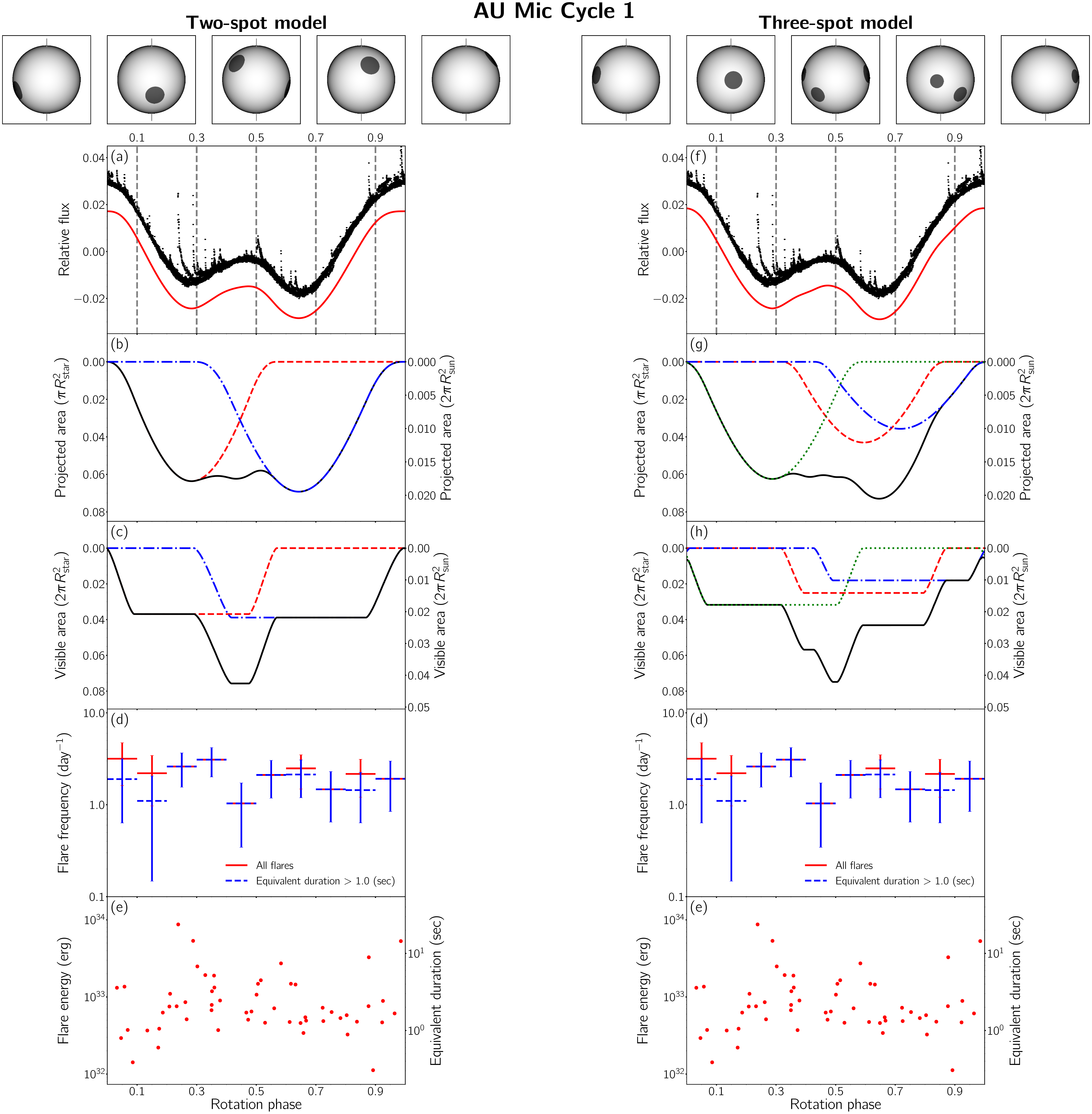} 
\caption{(Left) Visualized surface reproduced with each mode of posterior distributions for the rotation phases = 0.1, 0.3, 0.5, 0.7, and 0.9 (vertical dashed line). The light curve is folded with the period = 4.862 (day) from the day at the first peak = 1329.1 (BJD-2457000): (a) the phase-folded light curve of AU Mic Cycle 1 (black) and phase-folded reproduced one with the two-spot model (red); (b) the temporal variation of visible projected area of each spot (red, blue, and green) and the total area (black) relative to the stellar disk or solar hemisphere; (c) same as (b) but for the visible area relative to the stellar or solar hemispheres; (d) flare frequency per day in each bin for all flares (red) and flares with larger equivalent duration than $1.0$ (sec) (blue). The error bar equals a square root of (the number of flares in each bin + 1) per observation period; and (e) flare energy and the corresponding equivalent duration in each bin. (Right) (f, g, h) Same as the left but for the three-spot model. 
 \label{fig:phase1f}} 
\end{figure*}

\begin{figure*}[p]
\plotone{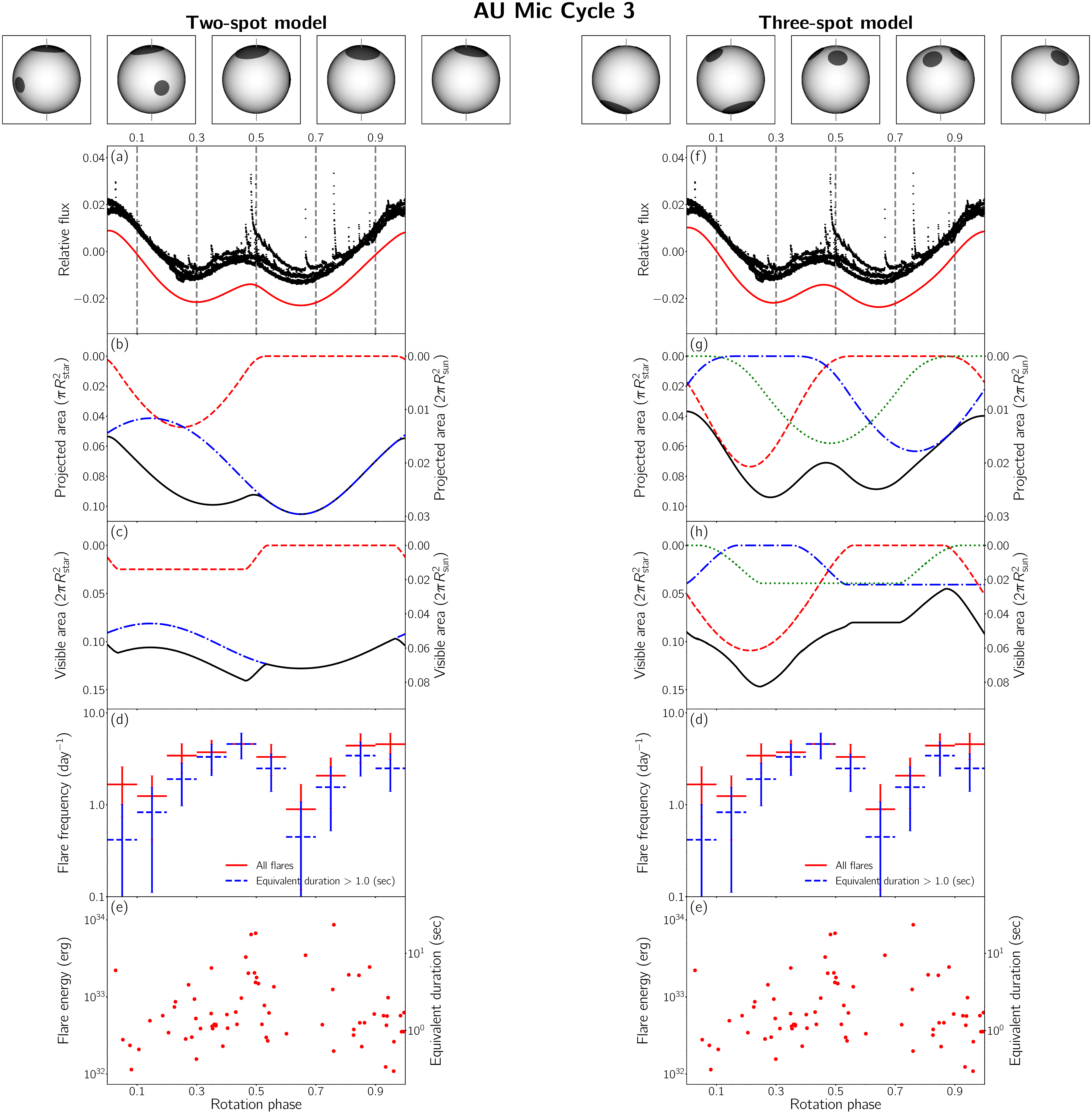} 
\caption{Same as Figure \ref{fig:phase1f} but for AU Mic Cycle 3 with the period = 4.829 (day) from the day at the first peak = 2040.1 (BJD-2457000).
\label{fig:phase1s}} 
\end{figure*}

\begin{figure*}[p]

\plotone{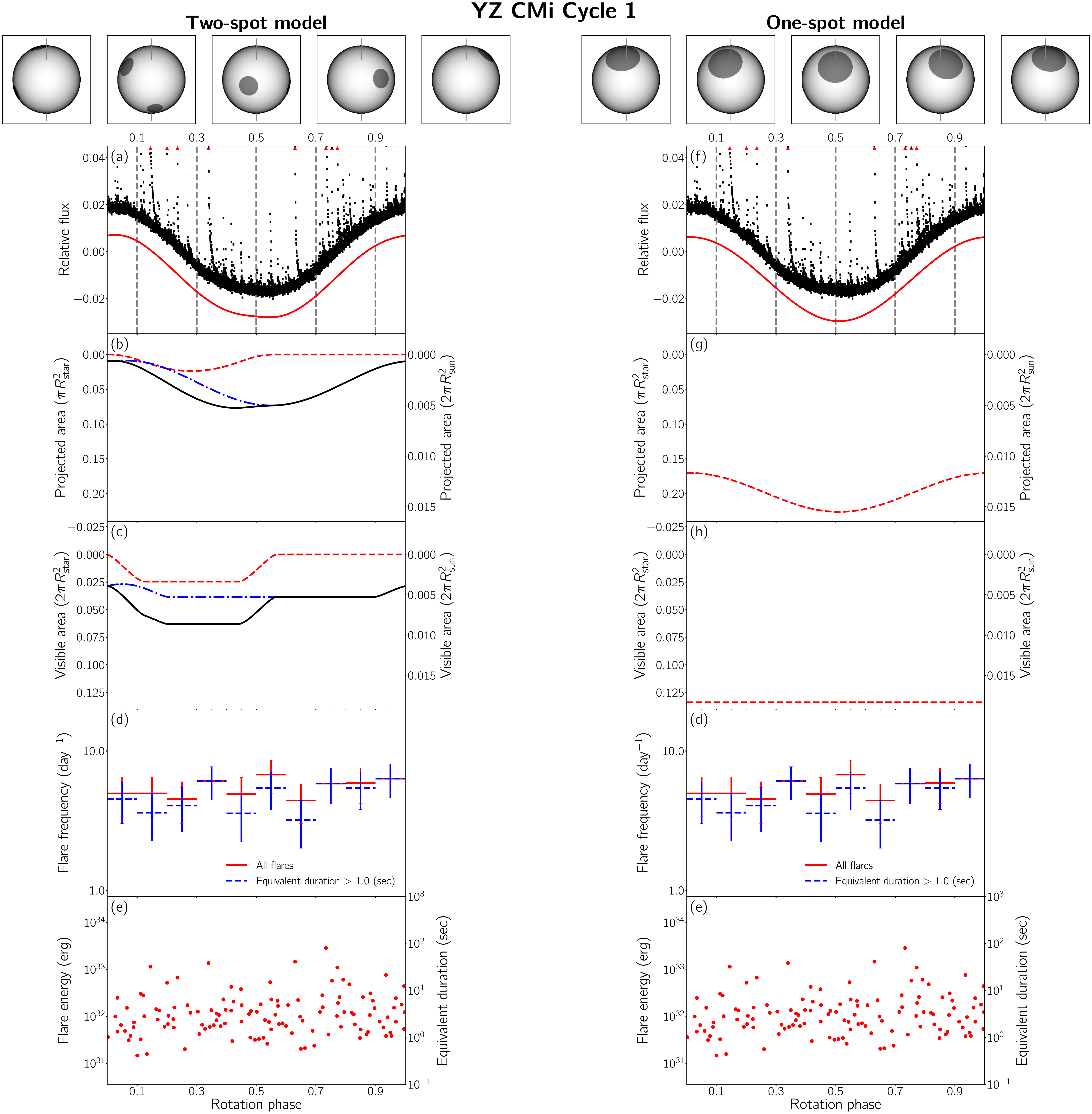} 

\caption{(Left) Visualized surface reproduced with each mode of posterior distributions for the rotation phases = 0.1, 0.3, 0.5, 0.7, and 0.9 (vertical dashed line). The light curve is folded with the period = 2.774 (day) from the day at the first peak = 1498.3 (BJD-2457000): (a) the phase-folded light curve of YZ CMi Cycle 1 (black) {with large flares marked as outliers (red triangle)} and phase-folded reproduced one with the two-spot model (red); (b) the temporal variation of visible projected area of each spot (red, blue, and green) and the total area (black) relative to the stellar disk or solar hemisphere; (c) same as (b) but for the visible area relative to the stellar or solar hemispheres; (d) flare frequency per day in each bin for all flares (red) and flares with larger equivalent duration than $1.0$ (sec) (blue). The error bar equals a square root of (the number of flares in each bin + 1) per observation period; and (e) flare energy and the corresponding equivalent duration in each bin. (Right) Same as the left but for the one-spot model. \label{fig:phase2f}} 
\end{figure*}

\begin{figure*}[p]
\plotone{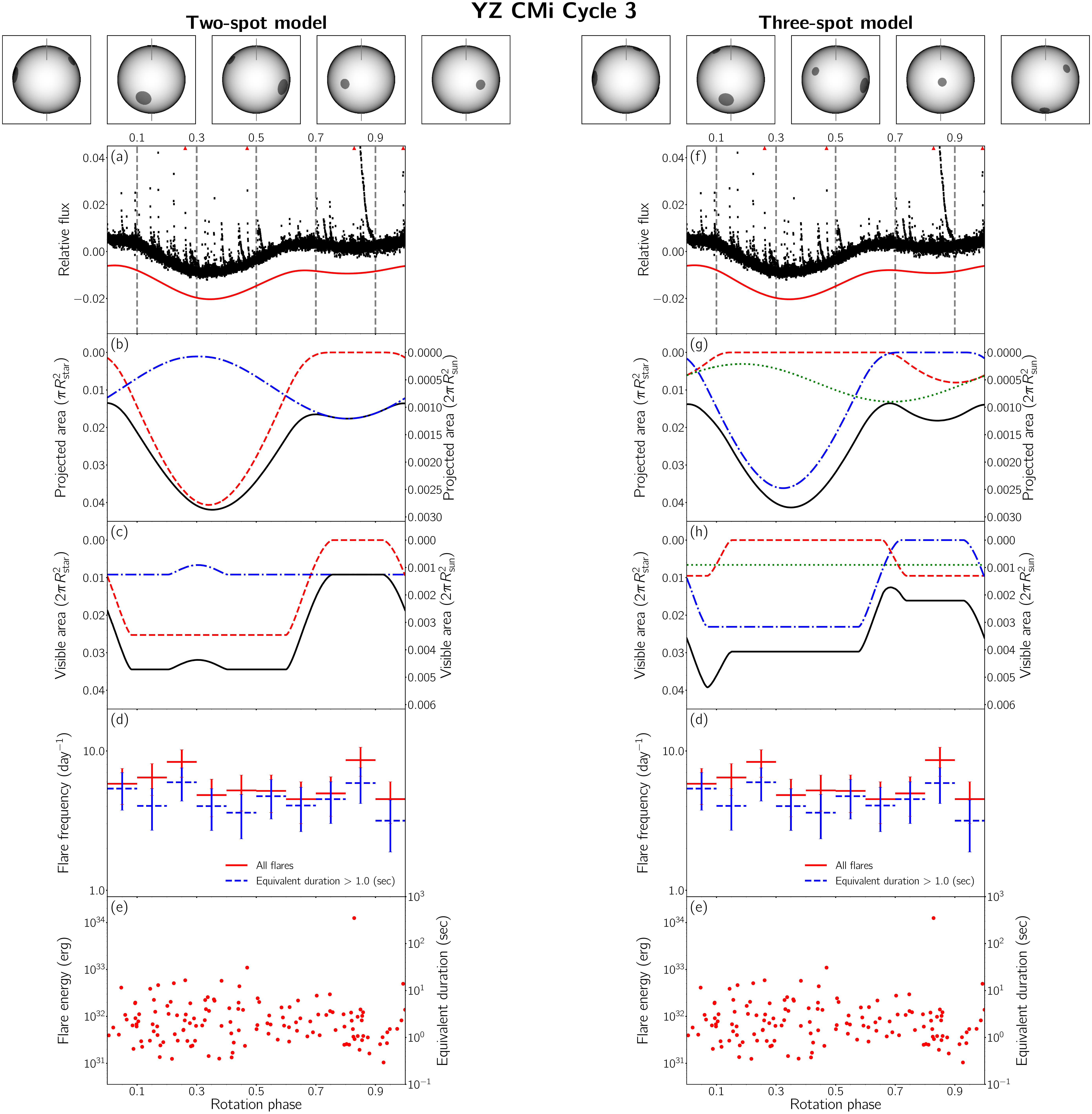} 
\caption{Same as Figure \ref{fig:phase2f} but for YZ CMi Cycle 3 with the period = 2.773 (day) from the day at the first peak = 2231.1 (BJD-2457000).\label{fig:phase2s}} 
\end{figure*}

\begin{figure*}[p]
\plotone{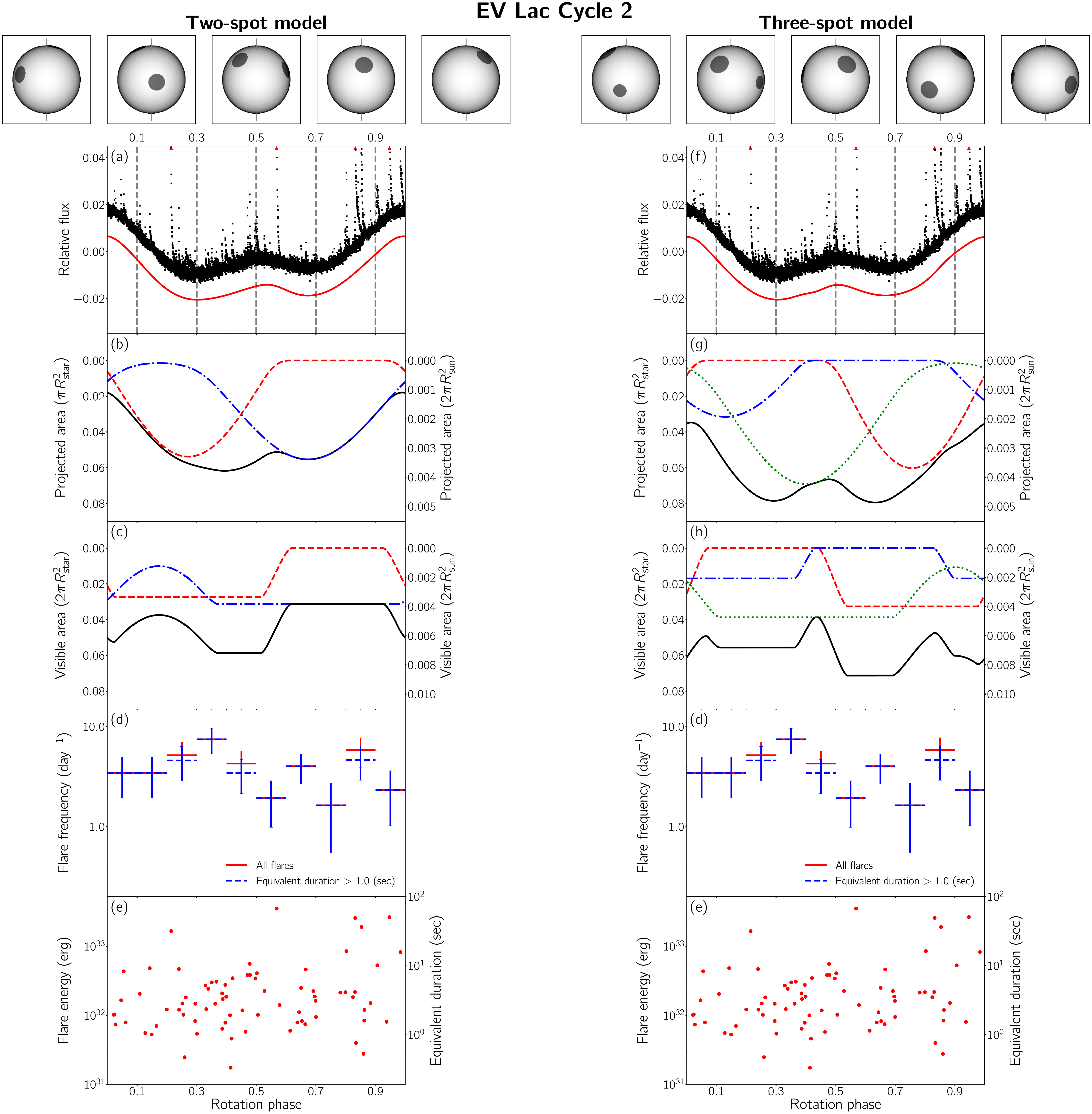} 
\caption{(Left) Visualized surface reproduced with each mode of posterior distributions for the rotation phases = 0.1, 0.3, 0.5, 0.7, and 0.9 (vertical dashed line). The light curve is folded with the period = 4.359 (day) from the day at the first peak = 1741.1 (BJD-2457000): (a) the phase-folded light curve of EV Lac Cycle 2 (black) {with large flares marked as outliers (red triangle)} and Phase-folded reproduced one with the two-spot model (red); (b) the temporal variation of visible projected area of each spot (red, blue, and green) and the total area (black) relative to the stellar disk or solar hemisphere; (c) Same as (b) but for the visible area relative to the stellar or solar hemispheres; (d) flare frequency per day in each bin for all flares (red) and flares with larger equivalent duration than $1.0$ (sec) (blue). The error bar equals a square root of (the number of flares in each bin + 1) per observation period; and (e) flare energy and the corresponding equivalent duration in each bin. (Right) Same as the left but for the three-spot model. \label{fig:phase3f}} 
\end{figure*}

\restartappendixnumbering

\begin{deluxetable*}{lccccc}[p]
\tablecaption{{EV Lac Cycle 2 additional case} \label{tb:para3s}}
\tabletypesize{\footnotesize}
\tablehead{
\colhead{Deduced Parameters} &  \colhead{Quadratic limb-darkening} & \colhead{Uniform} &\colhead{Normal}& \colhead{Gaussian process}& \colhead{Prior Distribution\tablenotemark{a}}  
}
\startdata
(Stellar parameters) &&&& &\\
1.  Equatorial period $P_{{\scriptsize \textrm{eq}}}$ (day) &$ 4.3482 ^{+ 0.0007}_{- 0.0010}$&$ 4.3554 ^{+ 0.0002}_{- 0.0004}$&$ 4.3559 ^{+ 0.0003}_{- 0.0003}$&$ 4.3518^{+ 0.0019}_{- 0.0106}$&${\cal U}_{\log} (4.0000,5.0000)$\\
2.   Degree of differential rotation $\kappa$ &$ 0.0147 ^{+ 0.0011}_{- 0.0010}$&$ 0.0037 ^{+ 0.0003}_{- 0.0002}$&$0.0032 ^{+ 0.0002}_{- 0.0001}$&$ 0.0120 ^{+ 0.0129}_{- 0.0026}$&${\cal U} (-0.2000,0.2000)$\\
(Spot parameters) &  & &   \\
(1st spot) &  &  &  &&\\
3.     Latitude $\Phi_1$  (deg) &$ 22.59 ^{+ 0.24}_{- 0.20}$&$ 25.70^{+ 0.31}_{- 0.39}$&$ 25.41 ^{+ 0.38}_{- 0.33}$&$ 21.11 ^{+ 1.92}_{- 1.52}$&{${\cal U} (-90.00,\Phi_2)$\tablenotemark{b}}\\
4.     Initial longitude $\Lambda_1$  (deg)&$ 50.30 ^{+ 0.13}_{- 0.10}$&$ 56.56 ^{+ 0.27 }_{- 0.20}$&$ 57.28 ^{+ 0.22}_{- 0.20}$&$ 50.30 ^{+ 0.73}_{- 0.71}$&${\cal U}  (-180.00,180.00)$\\
5.     Maximum radius $\alpha_{{\scriptsize \textrm{max,1}}}$  (deg)&$ 12.41 ^{+ 0.01}_{- 0.01}$&$ 24.35 ^{+ 0.21}_{- 0.17}$&$ 22.78 ^{+ 0.20}_{- 0.19}$&$ 12.46 ^{+ 0.05}_{- 0.04}$&${\cal U} (0.01,30.00)$\\
(2nd spot)&  &  & &&  \\
6.     Latitude $\Phi_2$  (deg) &$ 30.29 ^{+ 0.33}_{- 0.26}$&$ 46.14 ^{+ 0.65}_{- 0.60}$&$ 48.06 ^{+ 0.67}_{- 0.47}$&$ 29.16 ^{+ 1.64}_{- 1.89}$&{${\cal U} (\Phi_1,90.00)$\tablenotemark{b}}\\
7.     Initial longitude $\Lambda_2$ (deg) &$ -85.66 ^{+ 0.09}_{- 0.12}$&$ -86.97 ^{+ 0.17}_{- 0.14}$&$ -86.44 ^{+ 0.17}_{- 0.14}$&$ -85.56 ^{+ 0.79}_{- 0.72}$&${\cal U}  (-180.00,180.00)$\\
8.     Maximum radius $\alpha_{{\scriptsize \textrm{max,2}}}$ (deg)&$ 11.77 ^{+ 0.01}_{- 0.01}$
&$ 24.48 ^{+ 0.21}_{- 0.21}$&$ 23.14 ^{+ 0.22}_{- 0.22}$&$ 11.78 ^{+ 0.06}_{- 0.04}$&${\cal U}(0.01,30.00)$\\ \hline
(Additional parameters) & &  && &\\
9. Relative intensity $f_{\rm spot}$ &$0.48$ ({\it fixed})&$ 0.833 ^{+ 0.002}_{- 0.002}$&$ 0.811^{+ 0.003}_{- 0.003}$ &$0.48$ ({\it fixed})&${\cal U}  (0.00,2.00)$ / ${\cal TN} (0.48,0.01,0.00,2.00)$\\ 
10. GP amplitude $\ln \sigma_{\rm sys}$ &-&-&-&$ -6.5994 ^{+ 0.0181}_{- 0.0130}$&${\cal U}  (-10.0,0.0)$\\
11. GP exponential length $\ln \rho_{\rm sys}$ &-&-&- &$ -5.1250 ^{+ 0.0235}_{- 0.0198}$&${\cal U}  (-10.0,0.0)$\\ \hline
Average flux $F_{\rm ave}$ &  $0.854$ & $0.851$ & $0.851$ & $0.856$ & \\ \hline
{Logarithm of model} evidence $\log{\cal Z}$ &$-5995.666$& $-5654.106$ &$-5753.989$&$21813.240$&\\
\enddata
\tablenotetext{a}{
${\cal U_{\text{log}}}(a,b)= 1/(\theta \log(b/a))$, ${\cal U}(a,b)=1/(b-a)$, and ${\cal TN} (\mu, \sigma^2,a,b)$ represent bounded log uniform distribution (Jeffreys prior), bounded uniform distribution, and truncated normal distribution, defined in $a\leq \theta \leq b$, respectively.
}
\tablenotetext{b}{
We discern each spot by its latitude $\Phi_k$, not by its refereznce time $t_k$ in \citetalias{Ikuta20}, to improve the sampling efficiency of the PT.
}
\end{deluxetable*}

\section{Joint posterior distributions} \label{sec:appendix}

\begin{figure*}[p] 
\plotone{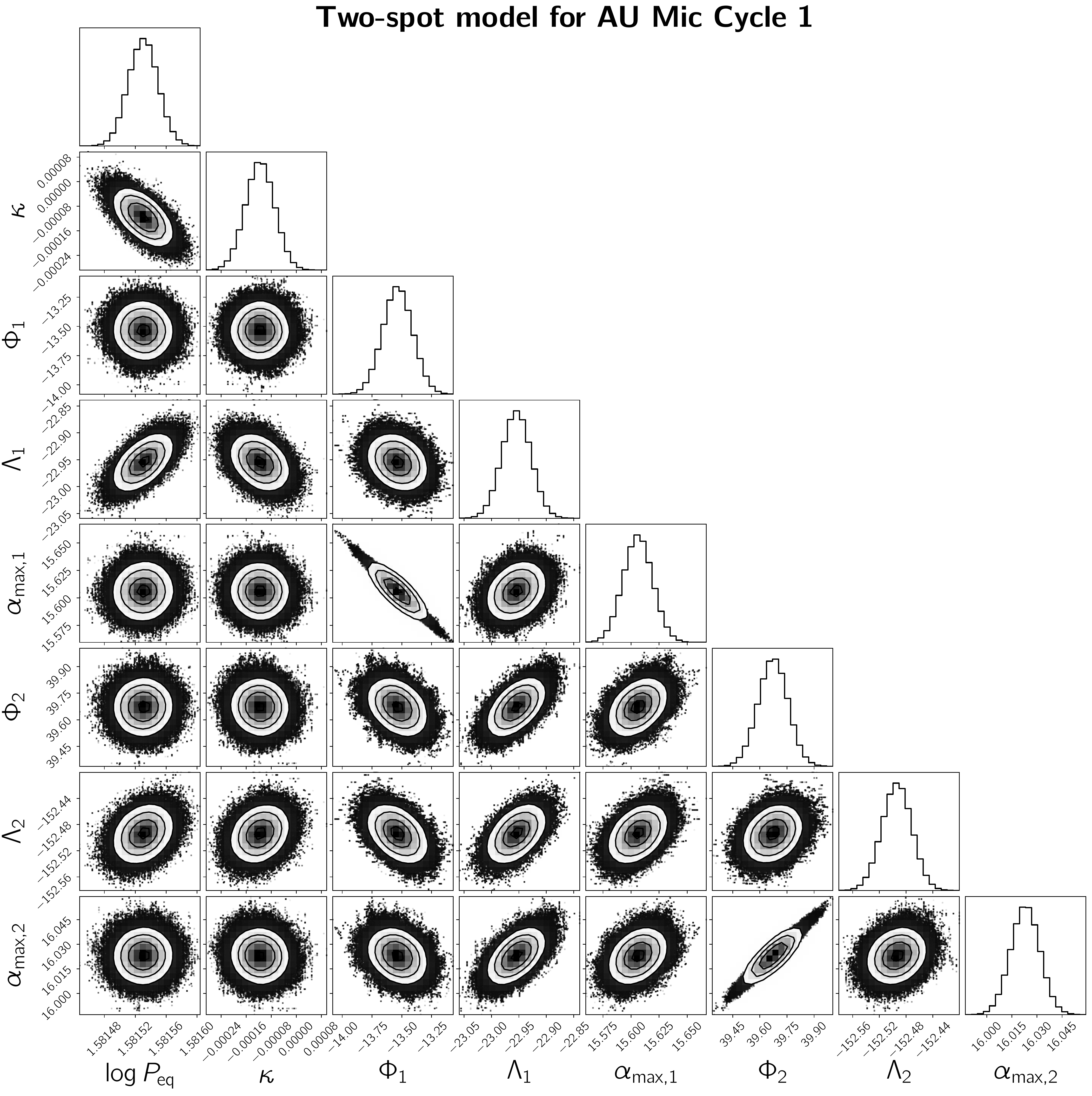}
\caption{The joint posterior distribution of parameters for the light curve of AU Mic Cycle 1 with the two-spot model. Each column represents the equatorial period $P_{\rm eq}$, degree of differential rotation $\kappa$, latitude $\Phi_k$, initial longitude $\Lambda_k$, and maximum radius $\alpha_{{\scriptsize \textrm{max}, k}}$, respectively. \label{fig:corner1}} 
\end{figure*}

\begin{figure*}[p] 
\plotone{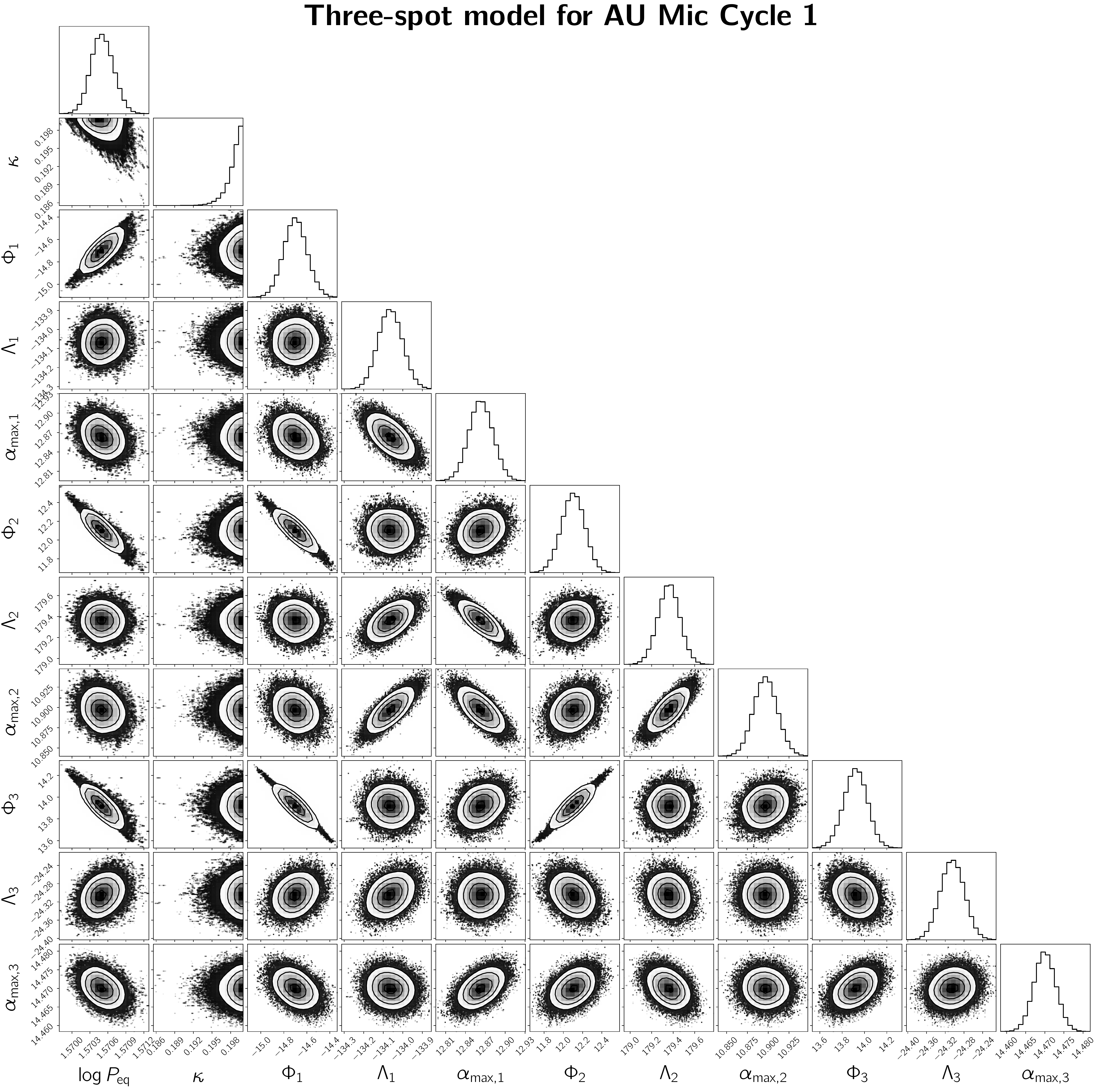}
\caption{Same as Figure \ref{fig:corner1} but for the three-spot model.
\label{fig:corner2}} 
\end{figure*}

\begin{figure*}[p] 
\plotone{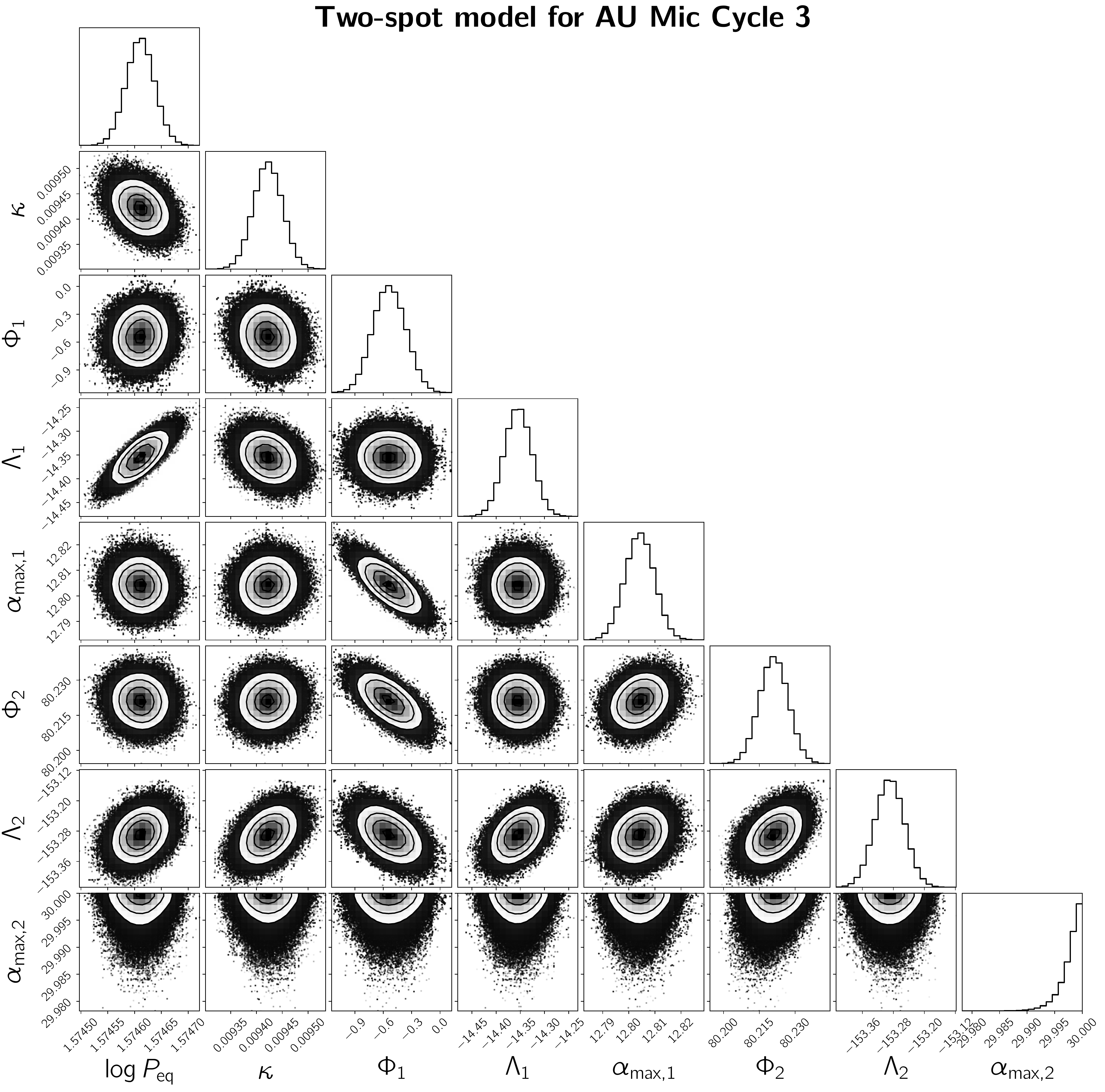}
\caption{Same as Figure \ref{fig:corner1} but for AU Mic Cycle 3. \label{fig:corner3}} 
\end{figure*}

\begin{figure*}[p] 
\plotone{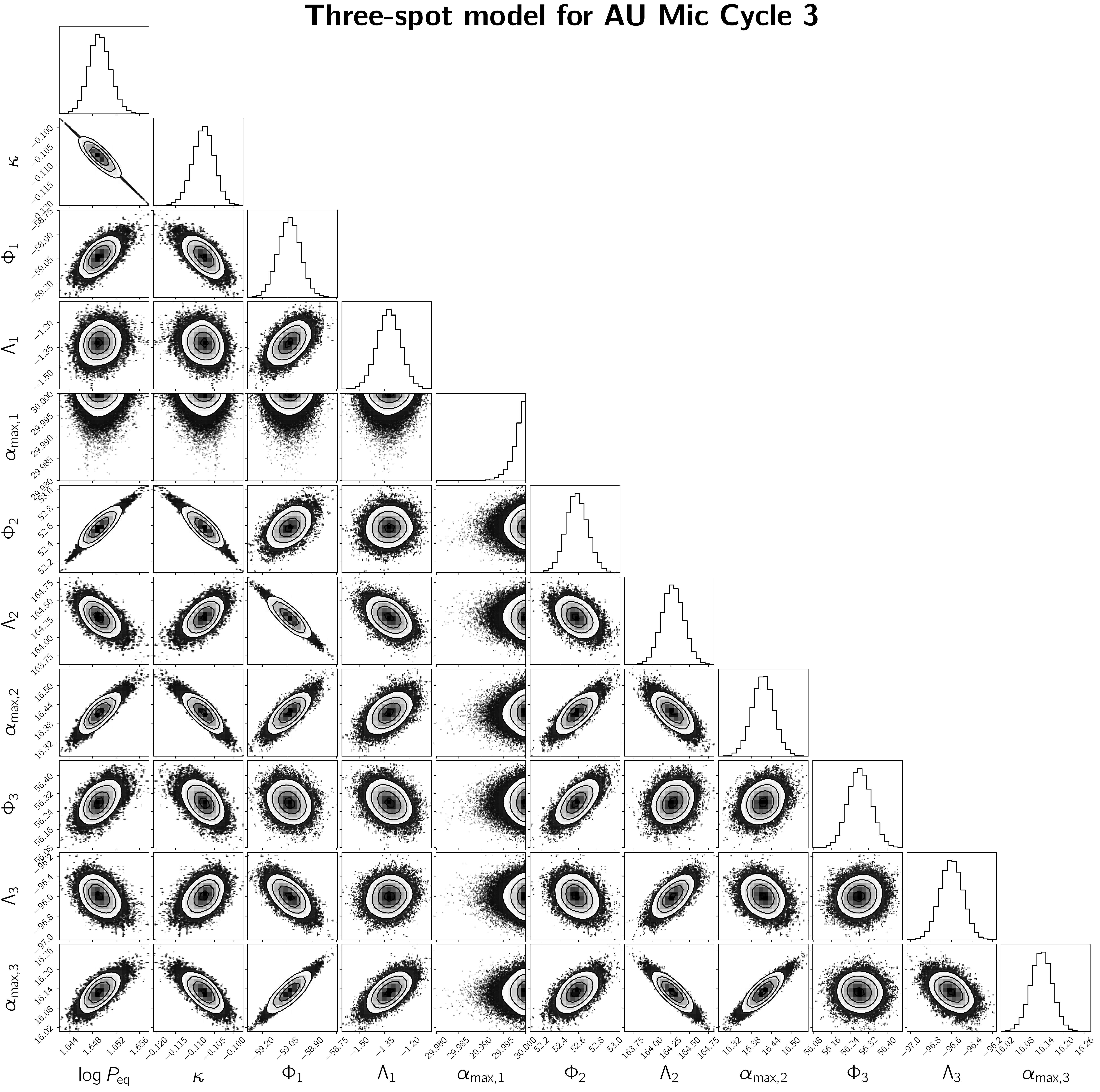}
\caption{Same as Figure \ref{fig:corner3} but for the three-spot model.
\label{fig:corner4}} 
\end{figure*}

\begin{figure*}[p] 
\plotone{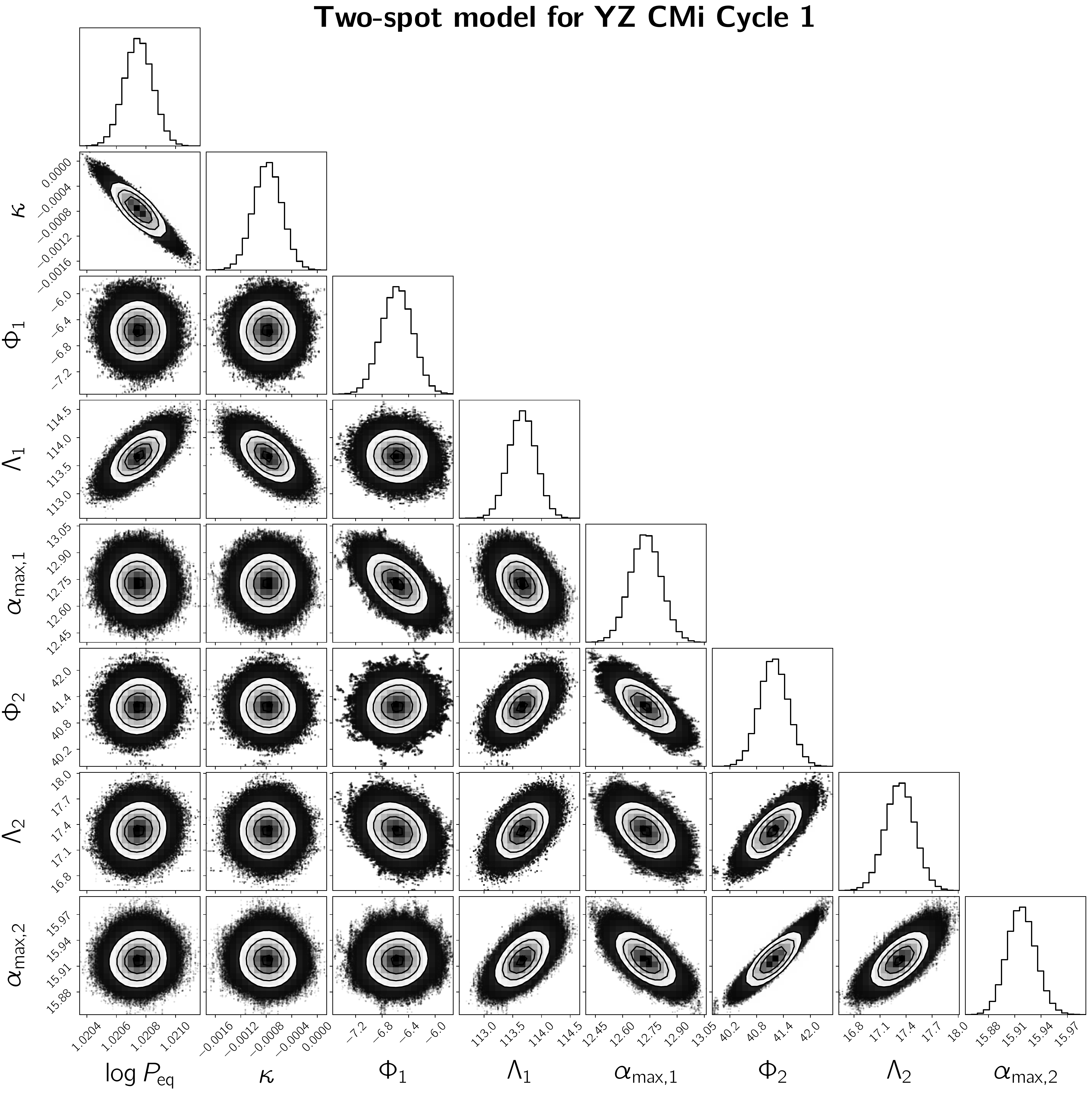}
\caption{The joint posterior distribution of parameters for the light curve of YZ CMi Cycle 1 with the two-spot model. Each column represents the equatorial period $P_{\rm eq}$, degree of differential rotation $\kappa$, latitude $\Phi_k$, initial longitude $\Lambda_k$, and maximum radius $\alpha_{{\scriptsize \textrm{max}, k}}$, respectively. \label{fig:corner5}} 
\end{figure*}

\begin{figure*}[p] 
\plotone{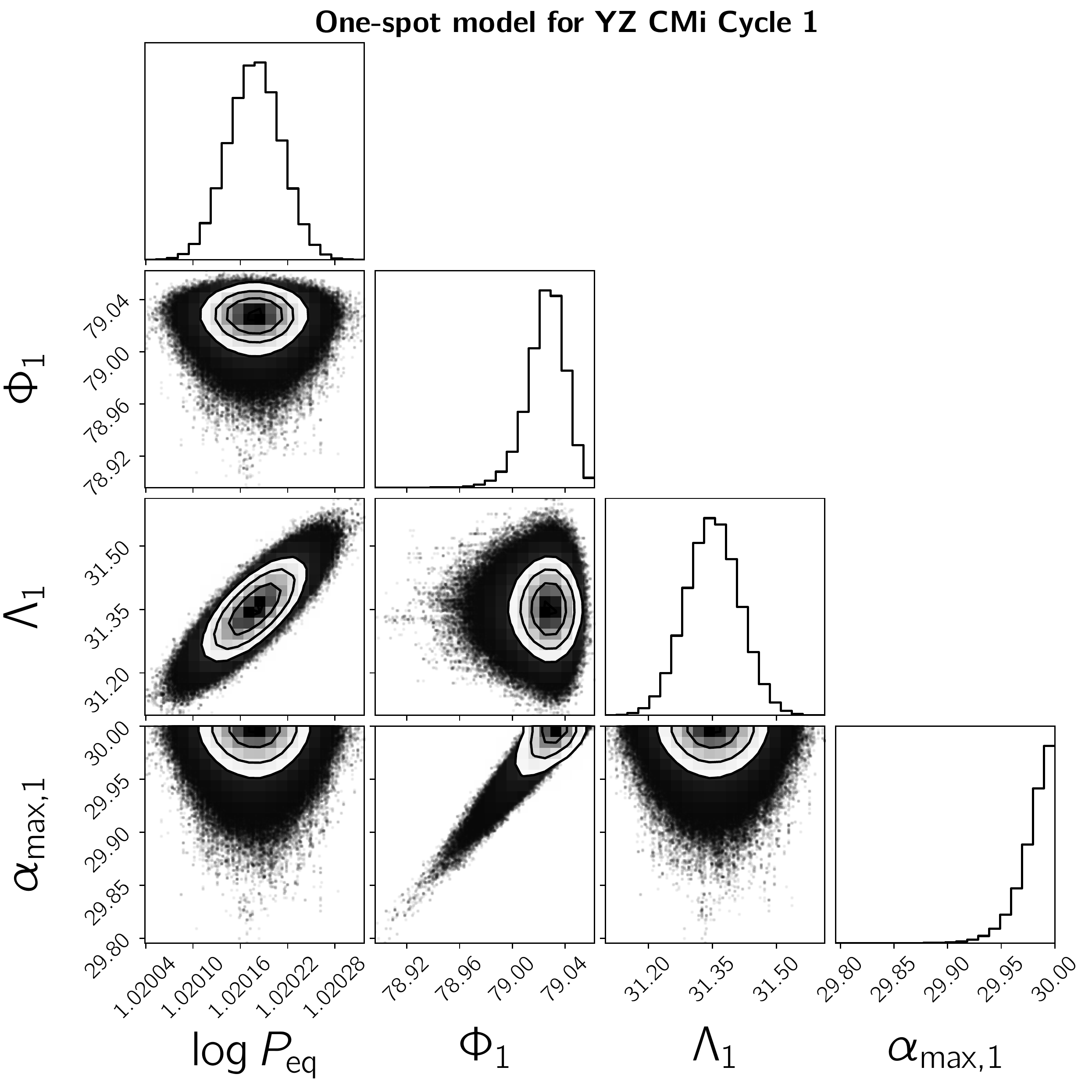}
\caption{Same as Figure \ref{fig:corner5} but for the one-spot model.
\label{fig:corner6}} 
\end{figure*}

\begin{figure*}[p] 
\plotone{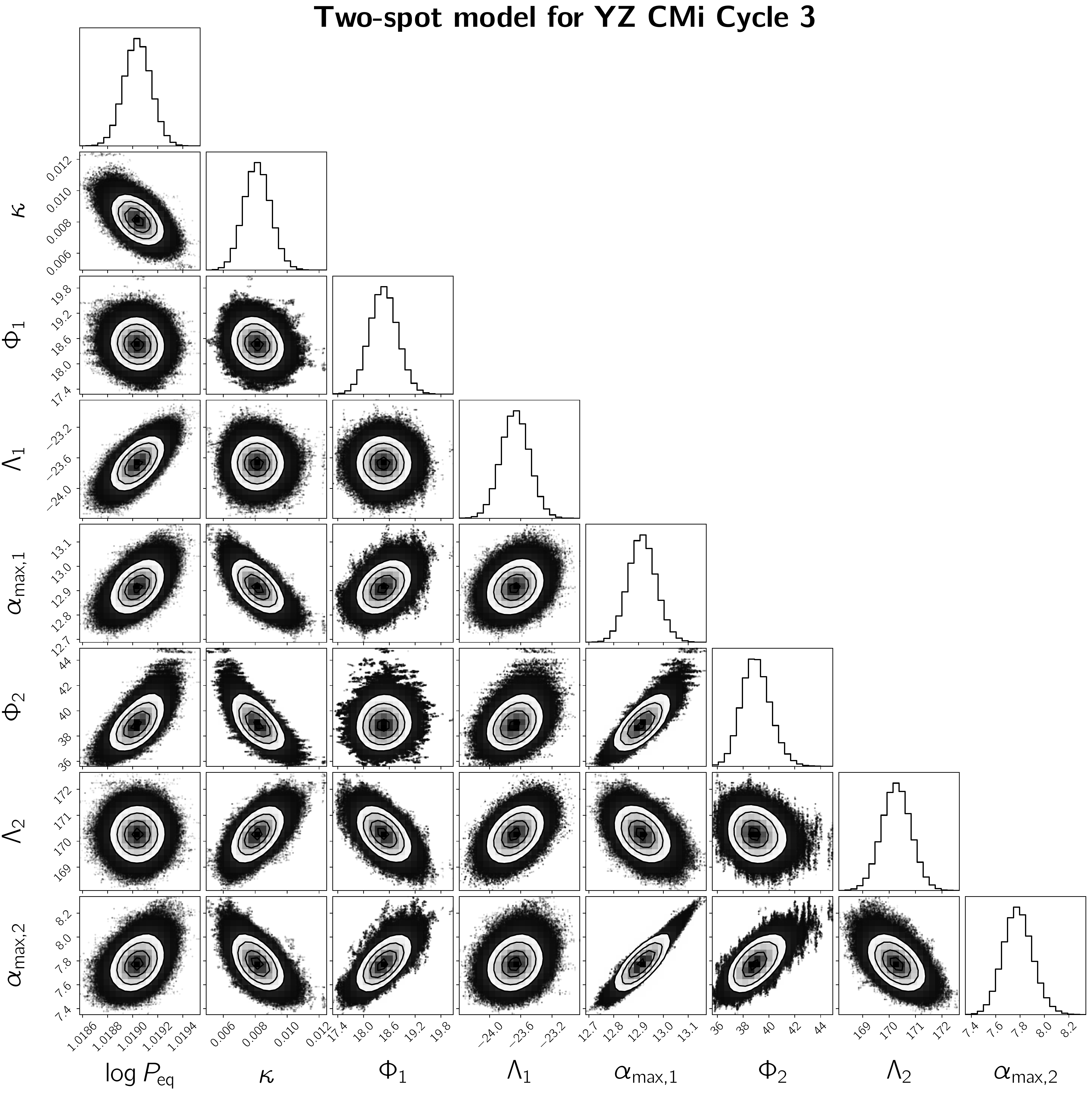}
\caption{Same as Figure \ref{fig:corner5} but for YZ CMi Cycle 3.
\label{fig:corner7}} 
\end{figure*}

\begin{figure*}[p] 
\plotone{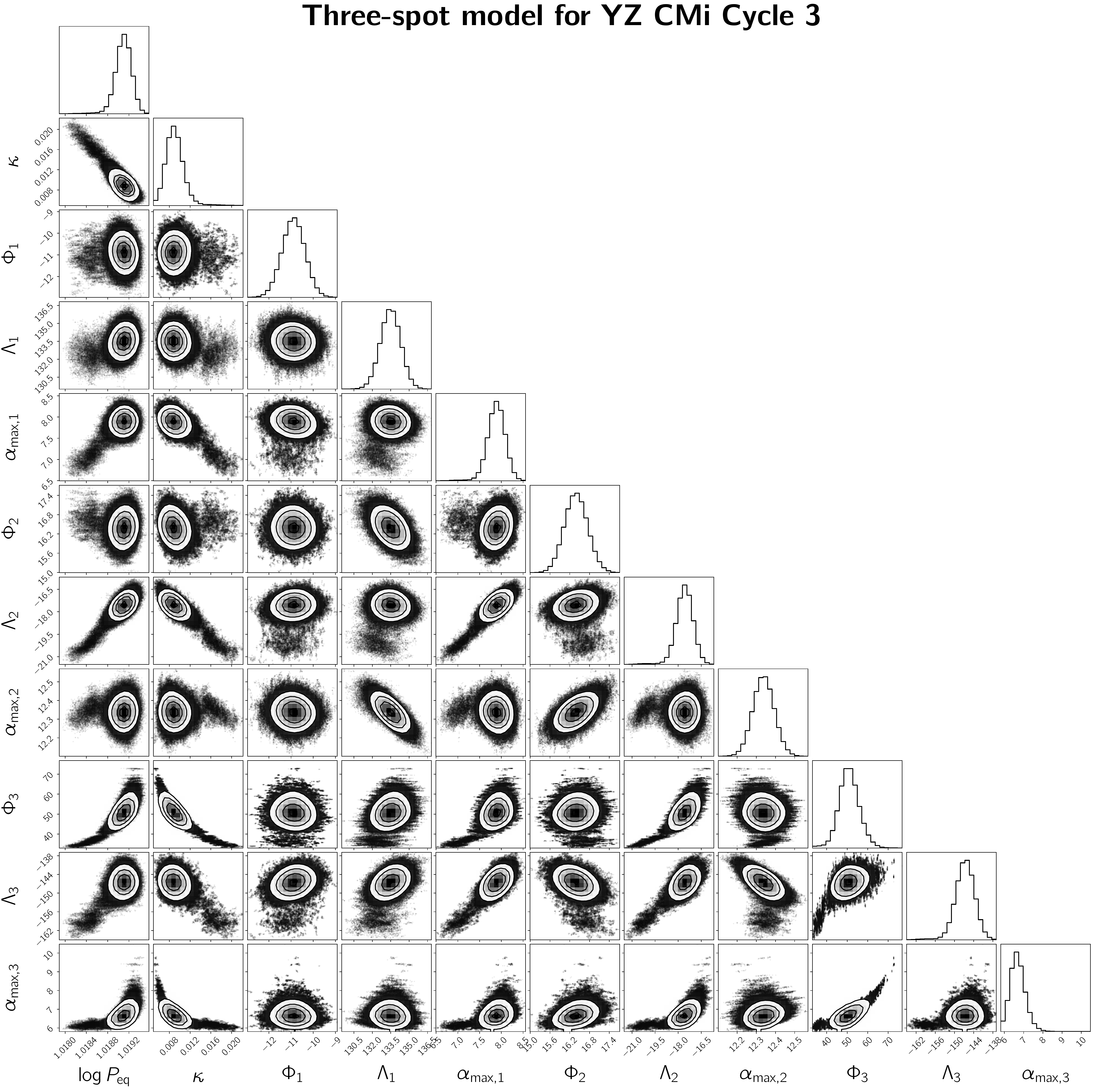}
\caption{Same as Figure \ref{fig:corner7} but for the three-spot model.
\label{fig:corner8}} 
\end{figure*}

\begin{figure*}[p] 
\plotone{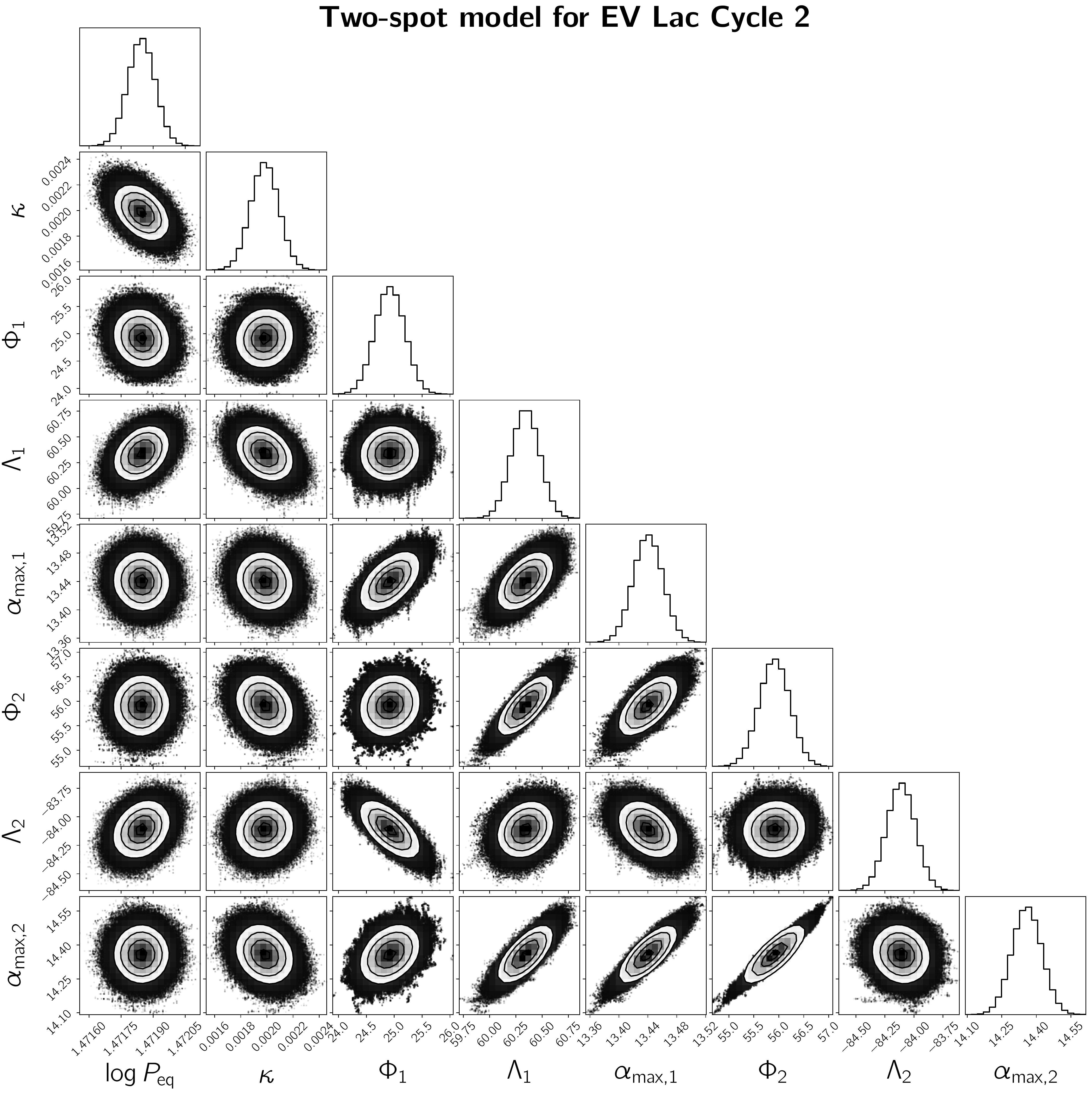}
\caption{The joint posterior distribution of parameters for the light curve of EV Lac Cycle 2 with the two-spot model. Each column represents the equatorial period $P_{\rm eq}$, degree of differential rotation $\kappa$, latitude $\Phi_k$, initial longitude $\Lambda_k$, and maximum radius $\alpha_{{\scriptsize \textrm{max}, k}}$, respectively. \label{fig:corner9}} 
\end{figure*}

\begin{figure*}[p] 
\plotone{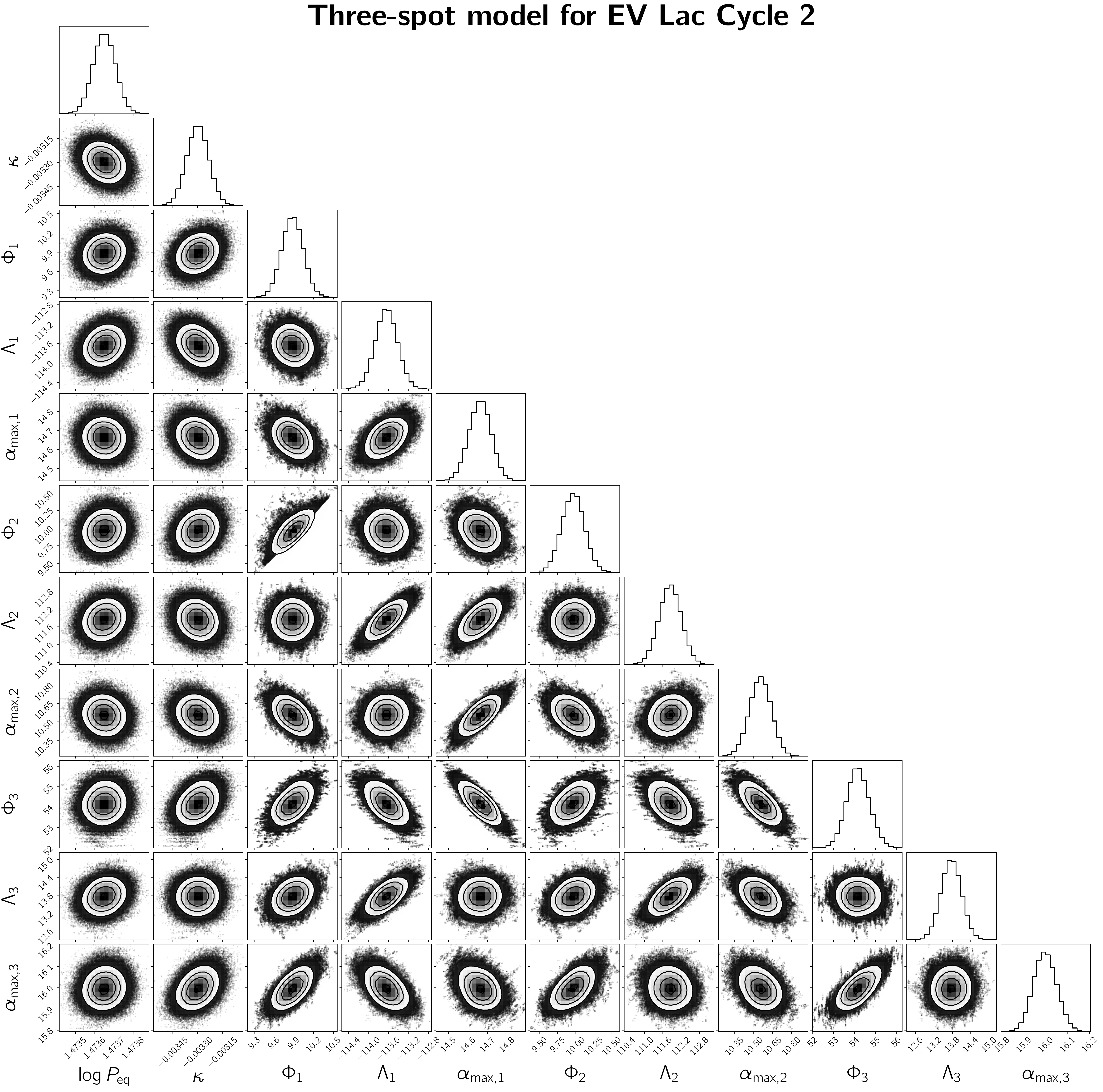}
\caption{Same as Figure \ref{fig:corner9} but for the three-spot model.
\label{fig:corner0}} 
\end{figure*}

The joint posterior distributions of 
the equatorial period $P_{\rm eq}$, degree of differential rotation $\kappa$, latitude $\Phi_k$, initial longitude $\Lambda_k$, and maximum radius $\alpha_{{\scriptsize \textrm{max}, k}}$,
are delineated in Figure \ref{fig:corner1}, \ref{fig:corner2}, \ref{fig:corner3}, \ref{fig:corner4}, \ref{fig:corner5}, \ref{fig:corner6}, \ref{fig:corner7}, \ref{fig:corner8}, \ref{fig:corner9}, and \ref{fig:corner0} for AU Mic Cycle 1 and 3, YZ CMi Cycle 1 and 3, and EV Lac Cycle 2, generated with \texttt{corner} \citep{2016JOSS....1...24F}. All figures show each of the distribution is an unimodal distribution.

\end{document}